\documentclass[a4paper,11pt]{article}
\pdfoutput=1 
\usepackage{jcappub} 
\usepackage{lineno}
\usepackage{graphicx}	
\usepackage{amsmath}	
\usepackage{hyphenat}
\usepackage{natbib}
\usepackage{booktabs,caption}
\usepackage{pdflscape}
\usepackage{hyperref}
\usepackage[export]{adjustbox}
\usepackage{tensor}
\usepackage{xcolor}
\usepackage{CJK}
\usepackage{aas_macros}
\usepackage[T1]{fontenc} 
\usepackage{xcolor}

\newcommand{\JP}[1]{\textcolor{black}{#1}}

\def\bfp{\mathbf{p}}

\arxivnumber{1234.56789} 
\title{\boldmath Compressed baryon acoustic oscillation analysis is robust to modified-gravity models}

\author[a,b]{Jiaming Pan}
\author[a,b]{Dragan Huterer}
\author[a,c]{Felipe Andrade-Oliveira}
\author[a,b]{Camille Avestruz}
\affiliation[a]{Department of Physics, University of Michigan, 450 Church Street, Ann Arbor, MI 48109, USA}
\affiliation[b]{Leinweber Center for Theoretical Physics, University of Michigan, Ann Arbor, MI 48109, USA}
\affiliation[c]{Physik-Institut, University of Zurich, Winterthurerstrasse 190, 8057 Zurich, Switzerland}

\emailAdd{jiamingp@umich.edu}

\abstract{We study the robustness of the baryon acoustic oscillation (BAO) analysis to the underlying cosmological model. We focus on testing 
 the standard BAO analysis that relies on the use of a template. These templates are constructed assuming a fixed fiducial cosmological model and used to extract the location of the acoustic peaks. Such ``compressed analysis'' had been shown to be unbiased when applied to the $\Lambda$CDM model and some of its extensions. However, it has not been known whether this type of analysis introduces biases in a wider range of cosmological models where the template may not fully capture relevant features in the BAO signal. In this study, we apply the compressed analysis to noiseless mock power spectra that are based on Horndeski models, a broad class of modified-gravity theories specified with eight additional free parameters. We study the precision and accuracy of the BAO peak-location extraction assuming DESI, DESI II, and MegaMapper survey specifications. We find that the bias in the extracted peak locations is negligible;  for example, it is less than 10\% of the statistical error for even the proposed future MegaMapper survey.  Our findings indicate that the compressed BAO analysis is remarkably robust to the underlying cosmological model.}


\begin{document}
\maketitle
\flushbottom



\section{Introduction}
Baryon acoustic oscillations (BAO) have by now established themselves as a main probe of cosmology, providing constraints on dark energy and the expansion history of the universe. The physics of the BAO has been well understood starting from the pioneering work by \cite{Peebles1970} and \cite{Sunyaev1970}: primordial sound waves in the baryon-photon fluid prior to recombination imprint a specific feature in the distribution of overdensities in the universe. This physical feature -- the sound horizon of about $r_{d} \approx 150$ Mpc in the standard cosmological model --- can be observed today in the distribution of galaxies as the scale at which there is a $\sim$10\% excess probability for clustering. The sound horizon allows for precise measurements of the angular diameter distance and the Hubble parameter at low redshifts ($z\sim 1$) where tracers of the large-scale structure --- galaxies and quasars --- are typically observed. 
The BAO feature was first detected and used to constrain cosmology nearly two decades ago \cite{Eisenstein2005ApJ,Cole2005}. Subsequent analyses have spearheaded an increasingly effective use of the BAO feature to constrain the cosmological parameters and models  (e.g.,~\ \cite{Anderson2012, Anderson2014,Alam2017, Alam2021}). Because the BAO features (in Fourier space, or a single feature in configuration space) reside in the linear-clustering regime, BAO are relatively free from systematic errors associated with nonlinear physics. BAO are thus a powerful tool to constrain different cosmological models of dark matter and dark energy. 

The standard BAO analysis --- the one that had been most commonly applied to data --- focuses on extracting the angular features corresponding to the sound horizon, while simply fitting out the broadband power spectrum. In other words, this kind of analysis fits a template that had been created assuming a fiducial cosmology. Galaxy clustering data are fitted to this template to extract the BAO feature(s), after marginalizing over many nuisance parameters which account for differences between the template and the measured broadband clustering. This standard analysis that makes use of a template is sometimes referred to as the \textit{compressed analysis} (since it compresses the clustering information into the transverse and radial location of the BAO peak), and we describe it in detail in Sec.~\ref{sec:method}. This procedure has been the basis for deriving cosmology from BAO measurements starting from the earliest analyses (e.g., \cite{Eisenstein2005ApJ,Kazin2010,Percival2010,Beutler2011,Beutler2017,Anderson2012,Anderson2014,Alam2017}). Alternatives to the standard analysis include the so-called "direct fit" (sometimes also called "full-shape modeling"), which fully models the broadband power spectrum including the BAO peak \cite{Ivanov2020,Philcox2020,Philcox2022,Troster2020}, as well as the \textit{ShapeFit} method, which is similar to the compressed analysis but includes a single additional parameter that extracts additional information about the slope of the power spectrum \cite{Brieden2021}. \JP{Additionally, one can also extract information from BAO signal using the "linear point" in the correlation function instead of the BAO peaks, which has the advantage of being less sensitive to nonlinear effects \cite{Anselmi2018}.}

It has been demonstrated that the compressed analysis is robust when one assumes the standard $\Lambda$CDM cosmological model \cite{Xu2012,Vargas2018}. In other words, the compressed analysis, which extracts the BAO peak location (or rather the relative location in the transverse direction, $\alpha_\perp$ and that in the radial direction, $\alpha_\parallel$; we will introduce these parameters in Sec.~\ref{sec:method}), recovers the true values of the cosmological parameters. This is not too surprising in the $\Lambda$CDM model, essentially because the compressed analysis is based on a template that had been constructed assuming $\Lambda$CDM.
However, it is possible that models beyond  $\Lambda$CDM add features to the power spectrum that cannot be well modeled by the template, and thus introduce unaccounted-for systematic errors that would bias the measured parameters. 

Relatively small deviations from $\Lambda$CDM (for example wCDM model, which adds as a free parameter the dark energy equation of state $w$) are expected to remain robust under standard compressed BAO analysis. This has been validated to some extent by previous studies that confirmed the flexibility and effectiveness of standard BAO analysis with different methodology choices and against data generated with different cosmological models. 
For instance, \cite{Carter2020} found that the extracted BAO peak-location parameters have negligible dependence on the assumed fiducial cosmologies, but their errors have a non-trivial increase when the fiducial cosmologies deviate from the test models. Similarly, \cite{Bernal2020} simulated the BAO  compressed analysis assuming that the cosmological models with modified perturbations before recombination, and found no significant shifts in the extracted cosmological parameters. 

However, the aforementioned studies have tested individual cosmological models, and no attempt to "sweep" through the much larger space of beyond-standard cosmological models has, to our knowledge, been attempted. The question remains therefore of precisely how robust is the compressed analysis in the presence of very general cosmological models, and how often (if at all) the standard BAO analysis fails under such models. We endeavor to answer this question, and reproduce the compressed BAO analysis algorithm, then apply it to a wide range of modified-gravity models. We will focus on Horndeski models (introduced and explained in Sec.~\ref{sec:horndeski}) for the following two reasons: (1) this is a very general class of modifications of gravity and is arguably well motivated to potentially explain the accelerating expansion of the universe, and (2) Horndeski models have been thoroughly studied in the literature, and in particular there exist publicly available Einstein-Boltzmann codes that produce the basic cosmological observables (like the primordial matter power spectrum) for an arbitrary model from this class.

The outline of the paper is as follows. We introduce the Horndeski models and describe their parameterization that we adopt along with theory parameter priors, in Sec.~\ref{sec:horndeski}. In Sec.~\ref{sec:method}, we describe the compressed BAO methodology, and specifically our implementation of it along with all relevant details and assumptions. In Sec.~\ref{sec:result}, we present the result of our tests for the robustness of BAO standard analysis in Horndeski models. We conclude and discuss other lines of current and future work in Sec.~\ref{sec:conclusion}.

\section{Horndeski models of  modified  gravity}
\label{sec:horndeski}

We now provide a concise overview of the Horndeski models. We present the selected parameters for the test models and discuss the impact of modified gravity on the resulting matter power spectra.

\cite{Horndeski1974} introduced models with the most general second-order Euler-Lagrange equations that can be obtained from the metric $g_{\mu\nu}$, the scalar field $\phi$, and their derivatives in four-dimensional space. Long after it was first proposed, the importance of the Horndeski framework was revisited and recognized by \cite{Charmousis2012} who reduced the original Lagrangian to a combination of four base Lagrangians. In this study, we follow the Effective Field Theory (EFT) approach \cite{Gleyzes2013,Gubitosi2013} that parameterizes the Horndeski models with a small number of free functions, which can be further reduced to a few parameters that control the cosmological background and perturbations.  

The action in unitary gauge for the EFT of dark energy can be written as the following (e.g., \cite{Gubitosi2013,Hu2014,Raveri2014}):

\begin{equation}
  \begin{aligned}
S_{\text{DE}} &= \int d^4 x \sqrt{-g} \bigg[ M_{\text{PL}}^2 [1+\Omega_{\text{MG}}(t)] \frac{R}{2} - \Lambda(t) - c(t) g^{00} \\[0.2cm]
&+\frac{M_{2}^{4}(t)}{2}(\delta g^{00})^2 - \bar{M_{1}}^3(t) \frac{1}{2} \delta g^{00} \delta K - \bar{M_{2}}^2(t) \frac{1}{2} (\delta K)^2  \\[0.2cm]
&- \bar{M_3}^2(t) \frac{1}{2} \delta \tensor{K}{_\nu^\mu} \delta \tensor{K}{_\mu^\nu}  + \hat{M}^2(t) \frac{1}{2} \delta g^{00} \delta R^{(3)}\\[0.2cm]
&+m_2(t) \partial_i g^{00}\partial^i g^{00} \bigg]+ S_{\text{m}}(g_{\mu\nu}, \Psi_{\text{m}}), 
  \label{eq:EFTDEaction}
  \end{aligned}
\end{equation}
where $M_{\text{PL}}$ is the Planck mass, $\delta g^{00}$ is defined as $g^{00}+1$, $\delta  \tensor{K}{_\mu^\nu}$ is the perturbation of the extrinsic curvature, $\delta K$ is its trace, $R$ is the Ricci scalar, $\delta R^{(3)}$ is the perturbation of the spatial component of the Ricci scalar, and $S_{\text{m}}(g_{\mu\nu}, \Psi_{\text{m}})$ is the action of matter field. There are a number of free functions here: $\Omega_{\text{MG}}(t)$, $\Lambda(t)$, $c(t)$, $M_2(t)$, $\bar{M_{1}}(t)$, $\bar{M_{2}}(t)$, $\bar{M_{3}}(t)$, $\hat{M}(t)$, and $m_2(t)$. The first three functions determine the background evolution; because $c(t)$ and $\Lambda(t)$ are subject to constraints from the energy density and pressure respectively in the Friedmann equations, the background evolution in modified gravity is controlled by the single function $\Omega_{\text{MG}}(t)$. In the literature, $\Omega_{\text{MG}}(t)$ is often referred to as $\Omega(t)$; we use $\Omega_{\text{MG}}(t)$ to avoid confusion with an energy-density parameter. The remaining free functions determine the evolution of perturbations.  

For convenience, we redefine the second-order free functions in a dimensionless form (see \cite{Bellini2014} for an alternative parameterization). The dimensionless functions are
\begin{equation}
    \begin{aligned}
          \gamma_1 &= \frac{M_{2}(t)^4}{m^2_0 H^2_0}, &
    \gamma_2 &= \frac{\bar{M}_{1}(t)^{3}}{m^2_0 H_0}, &
    \gamma_3 &= \frac{\bar{M}_{2}(t)^{2}}{m^2_0}, \\
    \gamma_4 &= \frac{\bar{M}_{3}(t)^2}{m^2_0}, &
    \gamma_5 &= \frac{\hat{M}(t)^{2}}{m^2_0}, &
    \gamma_6 &= \frac{m_{2}(t)^{2}}{m^2_0}.
    \end{aligned}
    \label{eq:gamma}
\end{equation}

In linearized Horndeski theory, the free functions governing the evolution of perturbations are subject to the following constraints:
\begin{equation}
    \begin{aligned}
    m_2 &= 0, \\
    \hat{M}^2 &= \frac{\bar{M_2}^2}{2} = -\frac{\bar{M_3}^2}{2}.
    \end{aligned}
    \label{eq:hornpri}
\end{equation}
The constraints in Eq.~(\ref{eq:hornpri}) imply $2\gamma_5 = \gamma_3 = -\gamma_4$ and $\gamma_6 = 0$. We adopt the following ansatz for the time-dependence of the remaining gammas:
\begin{equation}
\gamma_i(a) = \gamma_{i,0} \, a^{s,i},
\label{eq:gamma_parameteriz}
\end{equation}
since this functional form is simple yet reasonably flexible. Similarly, we choose $\Omega_{\text{MG}}(t)$ to have a form
\begin{equation}
\Omega_{\text{MG}}(a) = \Omega_{\text{MG,0}}(a) \, a^{\text{s,0}}.
\label{eq:OmegaMG_parameteriz}
\end{equation} 
Thus, there are eight free parameters  of the Horndeski models
\begin{equation}
    \bfp^{\rm Horn}= \{\Omega_{\text{MG,0}}, \gamma_{\text{1,0}}, \gamma_{\text{2,0}}, \gamma_{\text{3,0}}, s_{\text{0}}, s_{\text{1}}, s_{\text{2}}, s_{\text{3}} \}.
    \label{eq:pars}
\end{equation}
We adopt the \texttt{EFTCAMB} code \cite{Hu2014_2} to produce cosmological observables with Horndeski models described with the parameterization above.

The Horndeski parameters above specify the perturbations, but not the background. For the latter, we adopt the $\Lambda$CDM expansion history in a flat universe, with the single free parameter $\Omega_M=1-\Omega_\Lambda$. 

We now discuss the priors that we give to the Horndeski parameters; the priors are similar to (but not the same as) those adopted in \cite{Wen2023}.  The priors on $\gamma_{1,0}$ and $\gamma_{2,0}$ are chosen based on the preferred values from current cosmological data \cite{Frusciante2019}. The parameter $\gamma_3$ relates the speed of gravitational waves to the speed of light via 
\begin{equation}
\frac{c_T^2}{c^2} = 1 - \frac{\gamma_3(a)}{1 + \Omega_{\text{MG}(a)} + \gamma_3(a)},
\end{equation} 
where $c_T$ is the speed of gravitational waves. Here we choose $\gamma_{3,0} =0$; as gravitational waves propagate at the speed of light. In particular, the gravitational-wave event GW170817 ruled out all Horndeski models with $\gamma_{3,0} \neq 0$. Note that theories beyond general relativity, including the Horndeski class, allow non-luminal gravitational-wave speed at low energies. We set $\gamma_{3,0}=0$ in order to prevent models with non-luminal tensor speed at $z = 0$ (see \cite{Baker2022} for further discussion). All of these priors are summarized in Table \ref{tab:parameter}.  The parameterization of the cosmological parameters that control the background (and their associated priors) will be discussed below, in Sec.~\ref{sec:cosmo_models}.
 
Additionally, we impose physical stability conditions, mathematical stability conditions, and EFT additional conditions (see section IV F in \cite{Hu2014_2} for details). The physical stability, including both ghost and gradient stability conditions, ensures that the background evolution is stable (see Eqs.~(42)--(51) in \cite{Hu2014_2}). 
Ghost instability refers to a wrong sign of the kinetic term. Gradient instability is typically associated with a negative square of the sound speed, $c_s^2 < 0$ in the equations of motion of perturbations, leading to unbounded growth of small-scale perturbations. Mathematical stability conditions necessitate a well-defined $\pi$-field equation, the absence of fast exponential growth in the $\pi$-field perturbations, and a well-defined equation for tensor perturbations (see Eq.~(52) in \cite{Hu2014_2}). The mathematical stability conditions ensure that the perturbation in the dark section is stable (see Eqs.~(30)--(32) and ~(41)--(52) in \cite{Hu2014_2} for details of the physical and mathematical stability conditions effects on the parameters $\Omega_{\text{MG}}(a)$ and $\gamma_i(a)$). The EFT additional conditions require that  $w(a) \le -1/3$, which is already satisfied as we have fixed $w=-1$.

\begin{table}
	\centering
	\caption{The sampled ranges of Horndeski-model parameters that we considered to generate the mock power spectrum data. We assume a flat prior in all cases.}
	\label{tab:parameter}
	\begin{tabular}{lccr} 
		\hline
		Parameters & Lower bound & Upper bound \\
		\hline
		$\Omega_{\text{MG,0}}$& 0 & 0.1 \\
		$\gamma_{\text{1,0}}$ & 0 & 0.7 \\
		$\gamma_{\text{2,0}}$ & -1 & 0 \\
            $\gamma_{\text{3,0}}$ & fixed & fixed \\
            $s_{\text{0}}$ & 0 & 3 \\
            $s_{\text{1}}$ & -3 & 3 \\
            $s_{\text{2}}$ & 0 & 3 \\
            $s_{\text{3}}$ & N/A & N/A \\
		\hline
	\end{tabular}
\end{table}

\section{Simulating and measuring the BAO scale in Horndeski models}
\label{sec:method}

We now describe the methods that we used to test the bias of the BAO standard analysis. The procedure involves the following steps: 
\begin{enumerate}
\item We generate mock power spectra (and their multipoles), along with their corresponding covariance, based on each assumed galaxy survey and the underlying cosmological parameters. 
\item We fit the mock power spectrum multipoles using a template, thus jointly constraining about 15 cosmological and nuisance parameters. The parameters of our interest are the $\alpha$ parameters that describe the BAO location. 
\item We quantify the bias of the test model in the standard BAO analysis utilizing the best-fitted parameters and their confidence intervals. 
\end{enumerate}

The first two steps in this procedure are quantified in the rest of this Section. The fourth step constitutes our principal results, as outlined in Sec.~\ref{sec:result}.

\subsection{Cosmological model parameters}
\label{sec:cosmo_models}

To scan through a range of Horndeski cosmological models, we vary the Horndeski parameters given in Table \ref{tab:parameter}, as well as the cosmological parameters that specify the $\Lambda$CDM background. The Horndeski parameters $\bfp^{\rm Horn}$ are specified in Eq.~(\ref{eq:pars}), while the cosmological parameters are
\begin{equation}
    \bfp^{\rm cosmo}= \{h, \Omega_{\text{cdm}}h^2, \Omega_b h^2, A_s, n_s, \tau_{\text{reio}}  \},
    \label{eq:cosmo_pars}
\end{equation}
where $h$ is the Hubble constant in units of $100 \text{km}/\text{s}/Mpc$, $\Omega_{\text{cdm}}h^2$ and $\Omega_b h^2$ are respectively the physical cold dark matter and baryon energy densities, $A_s$ and $n_s$ are the amplitude and spectral index of primordial density fluctuations, and $\tau_{\text{reio}}$ is the optical depth of reionization.

\begin{figure}
      \centering
	\includegraphics[width=0.75\columnwidth]{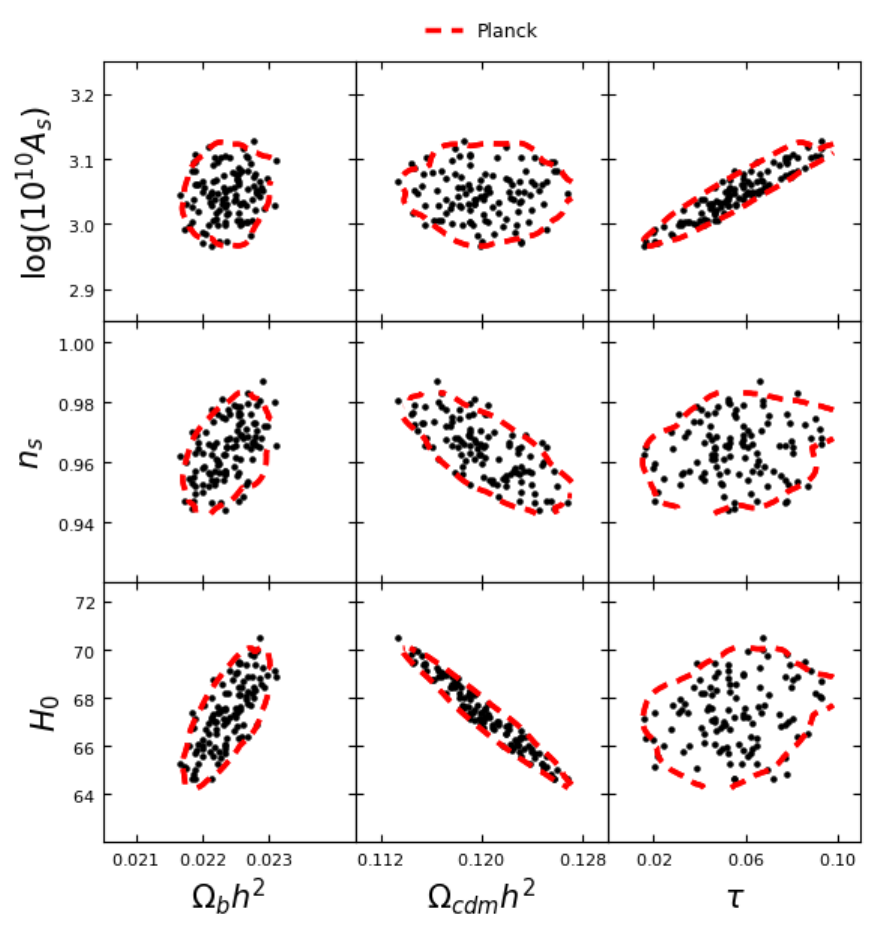}
    \caption{The parameters in the $\Lambda$CDM models for generating the mock power spectra data. The cosmological parameters were sampled within the 5$\sigma$ credible interval of the Planck posterior. These parameters are sampled simultaneously with the \texttt{EFTCAMB} Horndeski model parameters. The red contour shows the 5$\sigma$ credible (99.99994$\%$) ellipses for the Planck data with  TT+TE+EE+lowE constraint for comparison. }
    \label{fig:planck_sample}
\end{figure}

The Horndeski parameters are sampled from flat priors listed in Table \ref{tab:parameter}. In contrast, we choose a more complicated (correlated) prior for the base cosmological parameters from Eq.~(\ref{eq:cosmo_pars}) in order to make them in reasonably good agreement with respect to the current data; we do so since we do not wish to study models that are obviously ruled out. We choose the base cosmological parameters that generate $\Lambda$CDM  models that are within 5$\sigma$ confidence interval ($\Delta \chi^2 < 39.4$ in six-dimensional space from Eq.~(\ref{eq:cosmo_pars})) relative to the best-fit model from Planck
TTTEEE $+$ low E data \cite{Planck2020}\footnote{To be clear, we compute a covariance matrix for the parameters from the Planck chains and sample from the multivariate Gaussian distribution. Then, we apply an additional flat prior, requiring that each model be less than 5$\sigma$ ($\Delta \chi^2 < 39.4$) away from the Planck best fit. Because Horndeski parameters are not considered in this comparison with Planck (the perturbations are described by $\Lambda$CDM), it is likely that some models, when both $\Lambda$CDM background parameters and Horndeski (perturbation) parameters are eventually varied, are in a somewhat worse than 5$\sigma$ agreement with the current data.} We illustrate this in Fig.~\ref{fig:planck_sample}, where we  show the sampled $\Lambda$CDM parameters and the 5$\sigma$ contours of Planck data. We fix curvature to zero, and the neutrino density to $\Omega_{\nu} h^2 = 0.001$.


For each set of base cosmological and Horndeski parameters, we generate a matter power spectrum using in \texttt{EFTCAMB}. Subsequently, we utilize this derived matter power spectrum to compute $P(k,\mu)_{\text{data}}$ following the fitting template formulation that we describe in Sec.~\ref{sec:BAOanalysis} below.
We adopt values from the fiducial cosmology for the template nuisance parameters (that is, all parameters in the fit other than the relative location of the BAO peak, $\alpha_{\parallel}$ and $\alpha_{\perp}$; see also section~\ref{sec:BAOanalysis}).
We also need to specify a fixed set of cosmological parameters that describes the BAO template. We select the parameters that are close, but not identical, to the Planck best fit: $h = 0.6736, \Omega_{\text{cdm}}h^2 = 0.119, \Omega_b h^2 = 0.022, A_s = 2.1 \times 10^{-9}, n_s = 0.9649$, and $\tau_{\text{reio}} = 0.0544 .$

We next describe the procedure for calculating matter power spectra of Horndeski models. 


\subsection{Matter power spectra in Horndeski models}
\label{sec:matterPS}

We use \texttt{EFTCAMB} \cite{Hu2014_2} to generate matter power spectra predicted by Horndeski models. Note that the Horndeski parameters, listed in Eq.~(\ref{eq:pars}), only affect the perturbations and not the cosmological background (distances and volumes)\footnote{One might suspect that an exception to this is the Horndeski function $\Omega_{\text{MG}}(t)$ which appears to control the evolution of background. However, this function acts to rescale the Planck mass, as evident from Eq.~(\ref{eq:EFTDEaction}). Because the BAO peak positions as described in Eq.~(\ref{eq:ashif}) are described as distance ratios, the Planck mass cancels out and the BAO peak observable is not shifted under the change of $\Omega_{\text{MG}}(t)$. Note that the time evolution of Planck mass affects the anisotropic stress, but scaling the Planck mass by a constant value does not impact the observable quantities \cite{Bellini2014}.}, instead only affecting the shape of the BAO features in Fourier space. In contrast, the base cosmological parameters do shift the BAO peak; for example, the late-universe energy densities of dark matter and dark energy control the distance to the galaxies/quasars, and hence the angular extent of the sound horizon observed at the corresponding redshift. Therefore, each one of our models has a different BAO scale than that predicted by the fiducial $\Lambda$CDM model at that redshift. While the Horndeski parameter variations \textit{by themselves} do not shift the BAO, they change other features of $P(k)$ which are degenerate with those induced by varying the base $\Lambda$CDM parameters. Thus, the overall shift of the peak location in our models is more complex than that in vanilla $\Lambda$CDM.

\begin{figure}
      \centering
	\includegraphics[width=0.8\columnwidth]{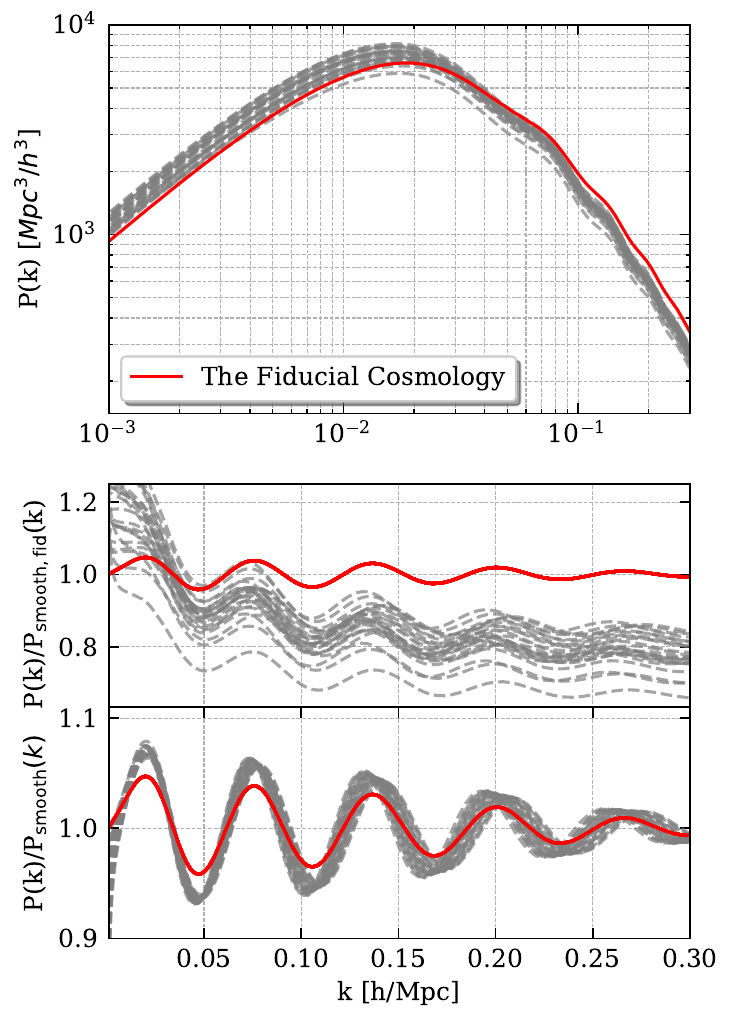}
    \caption{Twenty-four power spectra were randomly selected from our sampler, where both the Horndeski parameters and the base $\Lambda$CDM cosmological parameters were randomly sampled. The top panel shows the power spectra, the middle panel shows their ratios to the smooth fiducial power spectrum, $P(k)/P_{\text{smoooth, fid}}(k)$, while the lower panel shows the ratios relative to the smoothed portion of each corresponding spectrum, $O_{\text{lin}} = P(k)/P_{\text{smoooth}}(k)$. In each panel, the power spectrum evaluated in the fiducial cosmological model is plotted in red line. Note the non-negligible shifts in the BAO wiggle position. }
    \label{fig:pk_final}
\end{figure}


Fig.~\ref{fig:pk_final} illustrates the shifts in the location of BAO peaks for a representative sample of Horndeski models that includes variations of both the background (cosmological) and perturbation (Horndeski) parameters.
Note that the amplitude of the power spectrum and the locations of BAO peaks both vary in a complex way. We illustrate the $P(k)$ changes in more detail in Appendix~\ref{appendix:hornps}, where Fig.~\ref{fig:Omega s0} and Fig.~\ref{fig:EFTCAMB Paras} illustrate the change of matter power spectrum when the perturbation parameters alone are varied one at a time. We also present a comparison of the best-fitted parameters for the power spectrum in Horndeski models against those in the fiducial cosmology in Table~\ref{tab:bestparas}. We observe that the perturbation parameters alone induce some variation in the amplitude of the power spectrum, but not in the BAO location. At the same time, the base cosmological parameters do change the BAO peak locations as expected. Therefore, any given Horndeski model will have different BAO peak locations along with changes in the amplitude of the power spectrum that are potentially different than that expected from parameter variations in the $\Lambda$CDM model. This fact motivates our investigation, which is to see whether extraction of the BAO peak information in modified-gravity models that use a fixed template that is centered around $\Lambda$CDM can recover unbiased cosmological results.

\subsection{Template for the anisotropic power spectrum}
\label{sec:BAOanalysis}

We now \JP{review the standard BAO analysis} -- how to isolate and measure the BAO signal from our mock realizations \JP{in both isotropic and anisotropic cases}. Our analysis is anisotropic, i.e., separates the transverse and radial modes on the sky. Nevertheless, in the interest of pedagogy, we first review the isotropic analysis in order to introduce some key (and by now standard) tools.

To extract the BAO peak locations from data, it is economical to first fit a template to the power spectrum. The template assumes some fixed fiducial cosmological model and has a key feature of allowing freedom in the horizontal shift of the BAO features in k-space.  Specifically (and still assuming the isotropic BAO case for the moment), the BAO shift is controlled by the $\alpha$, defined as
\begin{equation}
\alpha \equiv \frac{D_{V}(z)}{r_d} \cdot \frac{r_{d,\text{fid}}}{D_{V,\text{fid}}(z)}.
\label{eq:ashif}
\end{equation}
Here we have defined a generalized distance \cite{Eisenstein2005ApJ}
\begin{equation}
  D_V(z) \equiv [cz(1+z)^2 D_A(z)^2 H^{-1}(z)]^{1/3},
	\label{eq:DV}
\end{equation} 
where $D_A(z)$ is the angular diameter distance, 
$H(z)$ is the Hubble parameter, and $r_d$ is the comoving sound horizon at the drag epoch. \JP{Here, $D_V$ is the distance that quantifies the average of the distances measured along, and perpendicular to, the line of sight to the observer. Moreover,} the subscript $``\text{fid}"$ refers to the corresponding values at the (fixed) fiducial cosmology, while $D_V$, $H(z)$, and $r_d$ are evaluated in the cosmological model that is being tested. 

The other parameters that enter the template, to which we will refer as the "nuisance parameters", also carry potentially useful cosmological information (about e.g.,\ the amplitude and shape of the primordial power spectrum), but are less robust than the $\alpha$ parameter as they are degenerate with systematic and astrophysical parameters, for example, the galaxy bias.  
We now introduce these remaining template parameters. We model the isotropic power spectrum following \cite{Anderson2014} 
\begin{equation}
   P(k) = P(k)_{\text{sm,iso}}[1+(O_{\text{lin}}(k/\alpha)-1)e^{-\frac{1}{2}k^2\Sigma^2_{\text{nl}}}],
	\label{eq:pkisotropic}
\end{equation}
where 
\begin{equation}
   P(k)_{\text{sm,iso}} = B_p^2P(k)_{\text{sm,lin}}+A_1k+A_2+\dfrac{A_3}{k}+\dfrac{A_4}{k^2}+\dfrac{A_5}{k^3},
	\label{eq:pksm}
\end{equation} 
and
\begin{equation}
   O_{\text{lin}}(k/\alpha) = \dfrac{P(k/\alpha)_{\text{lin}}}{P(k/\alpha)_{\text{sm,lin}}}.
\end{equation}
Here $B_p$, $\Sigma^2_{\text{nl}}$ and $A_i$ (with $i$ ranging from 1 to 5) are all nuisance parameters: $B_p$ accounts for potential large-scale bias, $A_i$ accounts for the possibility that $P(k)_{\text{sm}}$ does not match the actual data, and $\Sigma_{\text{nl}}$ characterizes the damping of BAO. Next, $P(k)_{\text{lin}}$ is the linear matter power spectrum, while $P(k)_{\text{sm,lin}}$ refers to the smooth part of the linear matter power spectrum, i.e.\ one without the BAO features (here we adopt a smoothing method in the configuration-space; see \ref{sec:procedure} for details).

Now we generalize the template to allow for anisotropy in the power spectrum (and the BAO peak locations). Instead of the single parameter $\alpha$, we now have $\alpha_{\parallel}$ and $\alpha_{\perp}$ that describe the BAO features in the parallel and transverse directions to the line-of-sight, respectively. They are defined as
\begin{equation}
\alpha_\perp\equiv \frac{D_{A}(z)}{r_d} \cdot \frac{r_{d,\text{fid}}}{D_{A,\text{fid}}(z)}, \\
\alpha_\parallel\equiv \frac{H_\text{fid}(z) r_{d,\text{fid}}}{H(z) r_d}.
\end{equation}
The fiducial cosmology is used to convert the measured redshift to the distance and establish the conversion factors between the fitting template and true values of $r_d$. The values of $\alpha_{\parallel}$ and $\alpha_{\perp}$ in the fiducial cosmology are unity. 

We next need to link the wavenumber utilized by \texttt{EFTCAMB}, $k \equiv |\mathbf{k}|$, and shift it to the wavenumber(s) that describe the anisotropic power spectrum in an arbitrary cosmological model. Starting with some $k$ that we provide to the numerical code, we first consider the cosine of the angle between this wavenumber and its projection along the line of sight, $\mu \equiv k_{\parallel}/k$; in this way, we get the wavenumber components that are respectively parallel and perpendicular to the line of sight, $k_{\parallel}$ and $k_\perp$. Next, we track how these two components scale to reflect the shift of the BAO peak in an arbitrary cosmological model. The "observed" wavenumbers in a given model are $k^{\text{aniso}}_{\parallel} = k_{\parallel}/\alpha_{\parallel}$ and $k^{\text{aniso}}_{\perp} = k_{\perp}/\alpha_{\perp}$, where $k_{\parallel}$ and $k_\perp$ are the values in the fiducial model. Finally, we can then express the coordinates of the anisotropic fitting in terms of the $k$ and $\mu$ using the following relationships: 
\begin{equation}
\begin{aligned}
k_{\text{aniso}} &= \frac{k}{\alpha_\perp} \left[1 + \mu^2 \left(\left(\frac{\alpha_{\perp}}{\alpha_{\parallel}}\right)^2 - 1\right)\right]^{1/2},\\[0.2cm]
\mu_{\text{aniso}} &= \frac{\mu}{\alpha_\parallel/\alpha_\perp} \left[1 + \mu^2 \left(\left(\frac{\alpha_{\perp}}{\alpha_{\parallel}}\right)^2 - 1\right)\right]^{-1/2}.    
\end{aligned}
\end{equation}

The fitting template needs to consider the effects of redshift-space distortions (RSD) and galaxy bias. First, the smooth component of the power spectrum takes the following form: 
\begin{equation}
P_{\text{sm}}(k, \mu) = B^2 (1+\beta\mu^2R(k))^2 F^2(k, \mu, \Sigma_{\text{FoG}}) P_{\text{sm,lin}}(k),
\label{eq:Pk_RSD}
\end{equation}
where the factor $B$ models the galaxy bias $b_g$ and the power spectrum amplitude variation. The term $(1+\beta\mu^2R(k))^2$ describes the effects of RSD at large scales \cite{Kaiser1987}. We have introduced the parameter $\beta = f/b_{g}$, where $f$ is the linear growth rate. 

\JP{The BAO oscillations are damped at high $k$ in the power spectrum due to non-linear gravitational collapse. The non-linear evolution also causes subpercent shift in the BAO peak location \cite{Eisenstein2007}. In order to revert the effects of non-linear evolution and thus enhance the statistical significance of the BAO signal, one can apply \textit{reconstruction}, a technique to shift the density field back to its original positions  \cite{Eisenstein2007_2}.} In that regard, the term $R(k)$ in Eq.~(\ref{eq:Pk_RSD}) that models the damping of non-linearities from RSDs on small scales can be modeled as \cite{Seo2016}
\begin{equation}
R(k) =
\left \{  
\begin{array}{cl}
1  & ({\rm pre-rec.}) \\[0.2cm]
1 - \exp\left(-(k\Sigma_{\text{sm}})^2/2\right)   & ({\rm post-rec}), \\[0.2cm]
\end{array} 
\right .
\end{equation}
where $\Sigma_{\text{sm}}$ is a smoothing scale, and where the first line refers to the pre-reconstruction case and the second line to post-reconstruction\footnote{\JP{The form of post-reconstruction $R(k)$ shown here corresponds to the \texttt{RecIso} convention used in the SDSS and (e)BOSS collaboration \cite{Padmanabhan2012, Alam2017, Alam2021}. There also exist alternative choices of the reconstruction formalism, like \texttt{RecSym} convention \cite{White2015}.}}. The term $F(k, \mu, \Sigma_{\text{FoG}})$ describes the Fingers of God effect \cite{Jackson:1971sky} which is the elongation of observed structures in redshift space along the line of sight, primarily caused by the peculiar velocities on small scales. This term is defined as (see Eq.~(27) in \cite{Peacock1994})  
\begin{equation}
F(k, \mu, \Sigma_{\text{FoG}}) = \frac{1}{1 + 0.5 (k\mu\Sigma_{\text{FoG}})^2},
\end{equation} 
where $\Sigma_{\text{FoG}}$ is the streaming scale. It is a free parameter within the fitting template, and we assume $\Sigma_{\text{FoG}} = 10$ Mpc/h in the fiducial cosmological model. 

The fitting template for the anisotropic matter power spectrum is  finally
\begin{multline}
    P(k_{\text{aniso}}, \mu_{\text{aniso}}) = P_{\text{sm}}(k_{\text{aniso}}, \mu_{\text{aniso}}) \times \biggl\{1 + (O_{\text{lin}}(k_{\text{aniso}}) - 1) \\
    \times \exp\left(-\frac{k_{\text{aniso}}^2}{2}\mu_{\text{aniso}}^2\Sigma_{\parallel}^2-\frac{k_{\text{aniso}}^2}{2}(1-\mu_{\text{aniso}}^2)\Sigma_{\perp}^2\right) \biggr\} + P_{\text{shot}},
    \label{eq:aniPS}
\end{multline} 
where $P_{\text{shot}} \equiv 1/\bar{n}$ and $\bar{n}$ is the mean number density of galaxies in the comoving volume. The shot noise results from the discrete distribution of galaxies and it is calculated assuming a Poisson distribution of the galaxies. The $\Sigma_{\perp}$ and $\Sigma_{\parallel}$ parameters model the transverse and
line-of-sight directions of the damping effects, respectively.

Due to the lack of availability of direct measurements of $\mu$ in typical observations, the standard BAO analysis utilizes the Legendre multipoles of the anisotropic power spectrum which marginalize over $\mu$. The monopole and the quadrupole moments of the fitting template of the anisotropic power spectrum take the following form:
\begin{equation}
P_0(k) = \JP{\left (\frac{r_{d,\text{fid}}}{r_{d}}\right )^3}\frac{1}{2\JP{\alpha_{\perp}^2\alpha_{\parallel}}} \int_{-1}^{1} P(k_{\text{aniso}}, \mu_{\text{aniso}}) d\mu + A_0(k)
\label{eq:mono}
\end{equation}
\begin{equation}
    P_2(k) = \JP{\left (\frac{r_{d,\text{fid}}}{r_{d}}\right )^3}\frac{5}{2\JP{\alpha_{\perp}^2\alpha_{\parallel}}} \int_{-1}^{1} P(k_{\text{aniso}}, \mu_{\text{aniso}})L_{2}(\mu) d\mu + A_2(k),
    \label{eq:qua}
\end{equation} 
where $L_2$ is the Legendre polynomial of the second order. \JP{The expression ($\frac{r_{d,\text{fid}}}{r_{d}})^3\frac{1}{\alpha_{\perp}^2\alpha_{\parallel}}$ accounts for the difference in the isotropic volume from the fiducial cosmology. Note that the factor $(\frac{r_{d,\text{fid}}}{r_{d}})^3$ is $k$-independent, and thus degenerate with (and can be subsumed in) the galaxy bias in the template-based analysis.} The polynomial terms $A_l$(k) are defined to be either one of these expressions
\begin{equation}
A_l(k)=
\left \{  
\begin{array}{ll}
a_{l,\text{1}} k^{-3} + a_{l,\text{2}} k^{-2} + a_{l,\text{3}} k^{-1} + a_{l,\text{4}} + a_{l,\text{5}}k  & \\[0.2cm]
a_{l,\text{1}} k^{-3} + a_{l,\text{2}} k^{-2} + a_{l,\text{3}} k^{-1} + a_{l,\text{4}} + a_{l,\text{5}}k^2  &
\end{array} 
\right .
\end{equation}
where the first line refers to the pre-reconstruction case and the second line to post-reconstruction. \JP{The form of $A_l$(k) for pre- and post-reconstruction is based on evaluating of the goodness of fit achieved by each term to the BOSS data \cite{Beutler2017}.}

Since we are investigating models of modified gravity, it is not appropriate to use the power spectrum where the peak locations were enhanced using information from galaxy velocities (the "post-reconstruction power spectrum"). This is because the reconstruction and its fitting template assume general relativity. Therefore, we study the monopole and quadrupole moments of the originally observed ("pre-reconstruction") power spectrum. In this case, there are 17 free parameters in the fitting template: 
\begin{align}\label{eq:templ_params}
\bfp^{\rm template} = \{&\alpha_{\parallel}, \alpha_{\perp}, B, \beta, \\[0.1cm]
&a_{0,1}, a_{0,2}, 
a_{0,3}, a_{0,4}, a_{0,5}, a_{2,1}, a_{2,2}, a_{2,3}, a_{2,4}, a_{2,5}, \notag\\[0.1cm]
&\Sigma_{\text{FoG}}, \Sigma_{\perp}, \Sigma_{\parallel}\}\notag.
\end{align}
In the fiducial cosmology, we choose the parameters of the template to take the following values:  $B = b_g$, $\beta = f/b_g$, $\Sigma_{\text{FoG}} = 10$ $\text{Mpc}/h$, $\Sigma_{\perp} = 4$ $\text{Mpc}/h$, $\Sigma_{\parallel} = 8$ $\text{Mpc}/h$. Moreover, all the coefficients of $A_l{(k)}$ are set to 0 and, as mentioned before, the fiducial values of the alphas are unity by construction ($\alpha_{\parallel} = \alpha_{\perp} = 1$). \JP{We choose the fiducial value of $\Sigma_{\text{FoG}}$ following \cite{Bernal2020}. The fiducial values for $\Sigma_{\perp}$ and $\Sigma_{\parallel}$ are motivated by \cite{Eisenstein2007}.} The value of $b_g$ depends on the survey and tracers under consideration and is presented in Table~\ref{tab:survey}. The template and its fiducial parameters are used in the calculation of the covariance matrix, which will be discussed in the next section.

\subsection{Fitting and extracting the BAO Signal}
\label{sec:fitting}

With the computed power spectrum in hand (Sec.~\ref{sec:matterPS}), and the description of how to model it (Sec.~\ref{sec:BAOanalysis}), it is fairly straightforward to extract the BAO feature. We do so by fitting the BAO model that is described by parameters in Eq.~(\ref{eq:templ_params}).

First, we clarify that we adopt noiseless data. That is, we use the theoretically predicted power spectrum multipole moments, with error bars as described below but without adding stochastic noise. We adopt the noiseless data to reduce the sample variance in our results. To be clear, we do expect additional \textit{statistical} error in the case with real data, but this statistical stochasticity operates independently of the biases caused by the insufficiently flexible template. It is precisely these latter effects that we wish to isolate and study. 

To perform the fit, we need to define our likelihood. It takes the following form 
\begin{equation}
    -2\log \mathcal{L} \propto [\mathbf{P}_{\text{th}} - \mathbf{P}_{\text{data}}]^T \mathbf{C}^{-1} [\mathbf{P}_{\text{th}} - \mathbf{P}_{\text{data}}],
\label{eq:likelihood}
\end{equation} 
where $\mathbf{P}_{\text{th}}$ is the concatenated vector of $k$-values of monopole and quadrupole moments of the power spectrum in the fitting template, and $\mathbf{P}_{\text{data}}$ is that of the mock power spectra.

Next, we need to specify $\mathbf{C}$, the covariance matrix of multipoles. We start with the matter power spectrum of a small bin size of $k$ and $\mu$, which can be approximated as 
\begin{equation}
\sigma^2(k, \mu) = \frac{8\pi^2P(k, \mu)^2}{k^2 \Delta k \Delta \mu V_{\text{eff}}},
\label{eq:errorbar}
\end{equation}
where $\Delta k$ is the width of $k$ bins and $\Delta \mu$ is  width of the $\mu$ bin. The error bar here includes both cosmic variance and shot noise; note that the latter is implicitly included given that it appears in the expression for the anisotropic power spectrum in Eq.~(\ref{eq:aniPS}). The effective volume \(V_{\text{eff}}\) is related to the measured physical volume by the equation:
\begin{equation}
V_{\text{eff}}(k,z) = \left(\frac{n(z)P(k,z)}{1+n(z)P(k,z)}\right)^2 V_{\text{survey}},
\end{equation}
where \(n(z)\) is the number density of galaxies. See Sec.~\ref{sec:result} for the chosen values of $k$ and $V_{\text{eff}}$ for the test models.

Assuming Gaussianity, the multipole covariance  is \cite{Grieb2016}
\begin{equation}
\begin{aligned}
        C_{\ell_1 \ell_2}(k_i, k_j) &= \delta_{ij}
        \frac{(2\ell_1 + 1)(2\ell_2 + 1)}{2} \\ 
        &\times 
        \int_{-1}^{1} d\mu \sigma^2(k_i, \mu)
        L_{\ell_1}(\mu) L_{\ell_2}(\mu).
        \label{eq:comulti} 
\end{aligned}
\end{equation}
This implicitly includes the sub-covariance matrices for each multipole and those between different multipoles. We used a fixed covariance matrix given a galaxy survey, evaluated in the fiducial $\Lambda$CDM model; this is likely to be sufficiently accurate and also reflects the procedure adopted in typical compressed BAO analyses.

We perform a global fit for all 17 parameters and effectively marginalize over 15 of them in order to obtain the posterior in the $(\alpha_\parallel, \alpha_\perp)$ plane. This approach stands in contrast to some other analyses that \textit{minimize} over the other template parameters to constrain the alphas (e.g., \cite{Bautista2021}). While the two approaches appear to give comparable results in practice for $\Lambda$CDM model, the marginalization that we adopt is likely to be more robust when a wider range of cosmological models is considered.

To constrain the parameters $\bfp^{\rm template}$,
we employ the Markov chain Monte Carlo (MCMC) algorithm (emcee, \cite{Foreman2013}), ensuring convergence by adhering to the Gelman-Rubin convergence criteria. Specifically, we set a threshold for $\hat{R}$ values at less than 1.001 for each parameter. The MCMC algorithm uses the likelihood function as in Eq.~(\ref{eq:likelihood}). We adopt flat priors on each of the free parameters in the analysis. The $\alpha_{\parallel}$ and $\alpha_{\perp}$ parameters are varied in the range $[0.8,1.2]$, while the rest of the parameters were assigned flat priors considerably wider than their final posterior values.

\begin{figure}
	\includegraphics[width=\columnwidth]{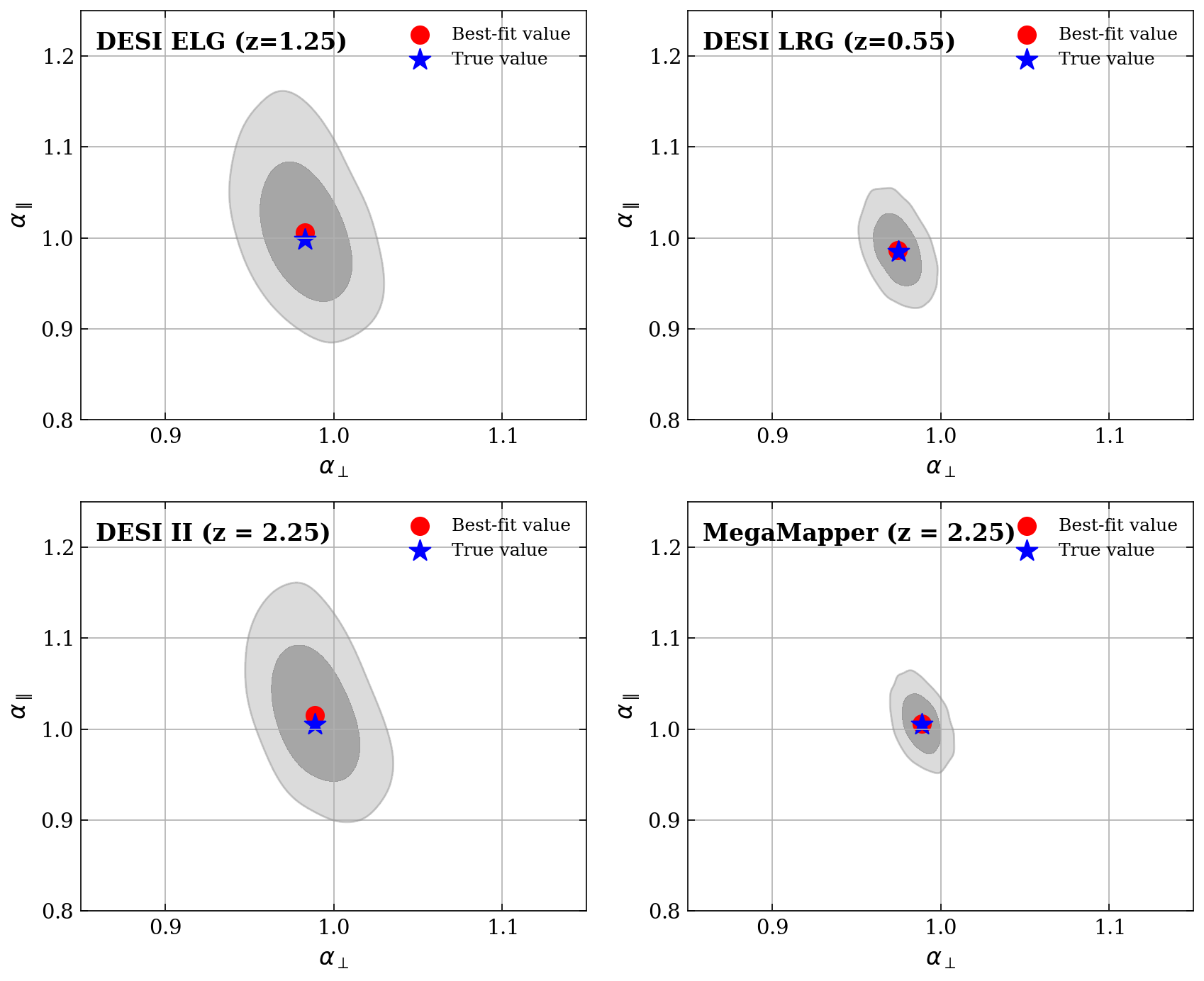}
    \caption{Likelihood contours in the $\alpha_{\perp}-\alpha_{\parallel}$ plane with 68\% and 95\% confidence levels. The red and blue dots indicate the maximum-posterior and true values, respectively. The plot shows an analysis of a single cosmological model using the compressed analysis (for several survey choices). For our results reported in Sec.~\ref{sec:result}, we repeat this procedure about 100 times for different cosmological models. We consider four distinct galaxy surveys and tracers: the DESI survey with LRG tracers at $z=0.55$ and ELG tracers at $z=1.25$, the DESI II survey with LBG tracers at $z=2.25$, and the MegaMapper Survey with LBG tracers at $z=2.25$.
     }
    \label{fig:alpha_plane}
\end{figure}

\subsection{Robustness Tests}
\label{sec:procedure}

Summarizing the procedure as outlined at the beginning of Sec.~\ref{sec:method}, we 
proceed as follows. For a given choice of Horndeski and background $\Lambda$CDM cosmological parameters, we first generate the power spectrum using \texttt{EFTCAMB} and compute its multipoles. 
Next, we fit these modified-gravity power spectra multipoles to the fitting template from Eqs.~(\ref{eq:mono}) and (\ref{eq:qua}). The fitting template is computed based on a power spectrum assuming fiducial cosmology (section~\ref{sec:fitting}) and has seventeen free parameters that are listed in Eq.~(\ref{eq:templ_params}).  Finally, we constrain all of these free parameters using the MCMC sampler. We repeat the procedure for 100 randomly sampled Horndeski models for each survey configuration described below. \JP{Furthermore, we investigate modified-gravity models where both background and perturbations significantly depart from $\Lambda$CDM.}

\begin{table}
    \centering
    \caption{Cosmological tracers in the adopted surveys. Here $z$ has information about the redshift slice adopted, $\Omega_{\text{Survey}}$ is the solid angle, $V_{\text{eff}}$ is the effective volume, $n_g$ is the average galaxy density, and $b_g$ is the galaxy bias that we adopted.
    }
    \label{tab:survey}
    \begin{tabular}{lccccc}
    \hline 
    \multicolumn{6}{|c|}{DESI SURVEY} \\
    \hline 
    Tracer & $z$ & $\Omega_{\text{Survey}}$ & $V_{\text{eff}}$ & $10^4n_g$  & $b_g$ \\
    & &$(\text{deg}^2)$ & $(\text{Gpc}/h)^3$ &(Mpc$^{-1}h)^3$ & \\
    \hline
    LRG & 0.55$\pm$0.15 &14,000& 4.24 & 4.84 & 2.16 \\
    ELG & 1.25$\pm$0.15 &14,000& 4.01 & 4.86 & 1.44 \\

    \hline 
   \multicolumn{6}{|c|}{DESI II SURVEY } \\
    \hline 
     LBG & 2.25$\pm$0.25 &10,000&7.76 & 2.77 & 2.12 \\
        \hline 
   \multicolumn{6}{|c|}{MegaMapper SURVEY } \\
    \hline 
     LBG & 2.25$\pm$0.25 &14,000& 10.87 & 7.90  & 2.12 \\
     \hline
    \end{tabular}
\end{table}

We simulate data of the quality expected from the Stage-IV experiment Dark Energy Spectroscopic Instrument (DESI), its planned extensions DESI-II, and the Stage-V experiment modeled on the proposed MegaMapper survey. For DESI, we have initially considered three different traces that have been commonly used in recent BAO analyses: luminous red galaxy (LRG), emission line galaxy (ELG), and quasars (QSO). We found that the QSO constraints are relatively weak given their lower number density, so we only carried out the DESI simulation with the LRG and ELG. We estimated the properties of these tracers based on the forecast in \cite{DESI2016,DESIvalidation2023}.\footnote{We estimate the galaxy bias for LRG, ELG, and QSO following section 6 in \cite{DESIvalidation2023}, with the linear growth factor normalized by $D(z = 0) = 1$. We determine the galaxy bias of LBG based on \cite{Bielby2013}, assuming the galaxy bias follows the form of $\frac{C}{D(z)}$ and C is a constant.}, and the tracer specifications are summarized in the top part of Table~\ref{tab:survey}. 
Here, we quote the mean redshift, the redshift bin width (equal to 0.3 for both DESI tracers considered), as well as the solid angle, effective volume, number density, and galaxy bias that we assumed.  In order to avoid the complexities of combining measurements from different redshift bins, for each tracer we only consider a single redshift bin.  

In Table~\ref{tab:survey} we also show our specifications for DESI  II and Megamapper. For these two future surveys, we have considered Lyman Break Galaxies (LBGs) as they represent the dominant population of objects expected to be observed  \cite{Schlegel2022}. The effective volume and galaxy density values of MegaMapper are based on forecasts on \cite{2023dAssignies}, and Table~\ref{tab:survey} also shows the other specifications assumed for DESI II and MegaMapper. Throughout, we employ $k$ values ranging from $0.0009$ to $1.10$ $h/\text{Mpc}$ to generate and analyze mock data.

\begin{table*}
    \centering
    \caption{ Summary of the statistics of $\Delta\alpha_{\parallel}$ and $\Delta\alpha_{\perp}$ --- differences between the measured and true values of the BAO peak-location parameters  --- based on 100 MCMC compressed analyses for different surveys and tracers. We show the mean and rms dispersion of each $\Delta\alpha$, along with the ratio of the rms dispersion in $\Delta\alpha$ to the statistical error in the corresponding $\alpha$, for each tracer type at specified redshifts in the DESI, DESI II, and MegaMapper surveys.}
    \begin{tabular}{lccccccc}
    \hline 
    \multicolumn{8}{|c|}{\textbf{DESI}} \\
    \hline 
    Tracer & $z$ & $\overline{\Delta\alpha_{\parallel}}$ & $\sigma(\Delta\alpha_{\parallel})$ &$\displaystyle\frac{\sigma(\Delta\alpha_{\parallel})}{\sigma(\alpha_{\parallel})}
$& $\overline{\Delta\alpha_{\perp}}$ & $\sigma(\Delta\alpha_{\perp})$ &$\displaystyle\frac{\sigma(\Delta\alpha_{\perp})}{\sigma(\alpha_{\perp})}$\\
    \hline
    LRG & 0.55$\pm$0.15 & 0.00196 & 0.00157&0.049 &-0.00025&   0.00052&0.035\\
    ELG & 1.25$\pm$0.15 & 0.00979 &0.00418&0.073 &0.00052&0.00111&0.051\\
    \hline 
   \multicolumn{8}{|c|}{\textbf{DESI II}} \\
    \hline 
    LBG & 2.25$\pm$0.25 & 0.00841& 0.00483&0.081&0.00091 &0.00175 &0.073\\
    \hline 
   \multicolumn{8}{|c|}{\textbf{MegaMapper}} \\
    \hline 
    LBG & 2.25$\pm$0.25 & 0.00104 & 0.00043&0.048&-0.00020&0.00134&0.044 \\
    \hline
    \end{tabular}
    \label{tab:delta_alpha}
\end{table*}



We have carried out a battery of tests to validate our approach and code. First, we were able to reproduce the results of \cite{Bernal2020}, where a very similar approach was adopted to study the robustness of several beyond-$\Lambda$CDM models to the compressed BAO analysis. We also studied the impact of different power-spectrum-smoothing methods to compute the power spectrum multiples. In particular, we compared the methodology utilized in the Barry code \cite{Hinton2017}\footnote{\cite{Hinton2017} compared different methods (polynomial regression, spline method, and the \cite{Eisenstein1998ApJ} method) to dewiggle power spectrum, and they found consistent results in these methods. Different methods to dewiggle the power spectrum therefore are not expected not affect the results of this study.} to that in \cite{Bernal2020}. The two methods agree very well; see Appendix~\ref{appendix:smoothb} for more details. We chose the configuration-space smoothing method in \cite{Bernal2020} as our fiducial approach.  

\JP{Additionally, we tested modified-gravity models that simultaneously vary both background and perturbations. 
To achieve simultaneous variations in the background and perturbations, we allowed freedom in the functions $\Omega_{\text{MG}}(t)$ and $\gamma_i(t)$.  We specified these two functions according to power-law expressions in Eqs.~(\ref{eq:gamma_parameteriz})-(\ref{eq:OmegaMG_parameteriz}), with corresponding amplitude and power-law coefficients sampled randomly from priors listed in Table~\ref{tab:parameter}. We further set the function $\Lambda(t)$ to zero. The background evolution in our modified-gravity models is then specified by $\Omega_{\text{MG}}(t)$ (see Eq.~(4) in \cite{Hu2014_2}), while the evolution of perturbations is controlled by it as well as by $\gamma_i(t)$. This framework allows us to explore the impacts on the compressed BAO analysis of modified-gravity models that significantly deviate from $\Lambda$CDM. The study of models where both background and perturbations vary from $\Lambda$CDM is not a primary goal of this paper, and these preliminary results are given in Appendix \ref{appendix:background}.}

Note there are two caveats for creating the mock power spectrum. 
First, the $\alpha_{\perp}$ and $\alpha_{\parallel}$ parameters in the mock power spectrum should be replaced by $q_{\perp}=\frac{D_{A}(z)}{D_{A,\text{fid}}(z)} $ and $q_{\parallel}=\frac{H_\text{fid}(z) }{H(z)}$. The reason is that the power spectrum has already been generated with the target true $r_d$, so there is no need to further scale the $r_d$ from its fiducial value. Second, the units for the power spectrum $P(k,\mu)_{\text{data}}$ need to include the same scaled Hubble constant $h$ as that used in the fitting template (that is, the fiducial value of $h$ adopted in the template). Only then can the correct definitions for $\alpha_{\perp}$ and $\alpha_{\parallel}$  be recovered from the fitting template. This is because there are no free parameters available to rescale $h$ between the units of the mock power spectrum and those employed in the template.

\section{Results}
\label{sec:result}

We now present our main results. We are interested in the constraints on $\alpha_{\parallel}$ and $\alpha_{\perp}$, marginalized over the other 15 template parameters. Note that we are not particularly interested in the statistical error of the $\alpha$ parameters itself, given that it is dependent on our choice of the redshift bin along with all other survey specifications (which may end up being different in reality from what we assume here). Rather, we focus on the biases in the best-fit value of the alphas \textit{relative} to the size of their statistical error.

In Fig.~\ref{fig:alpha_plane} we present the $\alpha_{\parallel}-\alpha_{\perp}$ log-likelihood contour for \textit{one} example randomly chosen modified-gravity model\footnote{This model has parameter values of $\gamma_1(a) = 0.430\, a^{-1.940}$, $\gamma_2(a) = -0.792\, a^{0.649}$, $\gamma_3(a) = 0$, and $\Omega_{\text{MG}} = 0.056\,a^{0.309}$.} and for several survey configurations. The blue star indicates the true values of the alphas (which are quite close to unity for this particular Horndeski model), the red circle shows the best-fit value from our procedure described in Sec.~\ref{sec:method}, and the grey contours show the 68\% and 95\% credible contours for our fit, marginalized over all of the fit parameters. This figure already previews our key result, which is that the recovered alphas are in excellent agreement with their true values. In other words, the bias in the recovered alphas relative to their statistical values is well below 1 $\sigma$. This is true for both tracers of DESI, as well as DESI II and MegaMapper (the latter of which has a small forecasted error in the alphas even in a single redshift bin that we assumed). We therefore infer, from just one model for the moment, that the compression appears to be robust when applied to Horndeski models.

\begin{figure}
     \includegraphics[width=\linewidth]{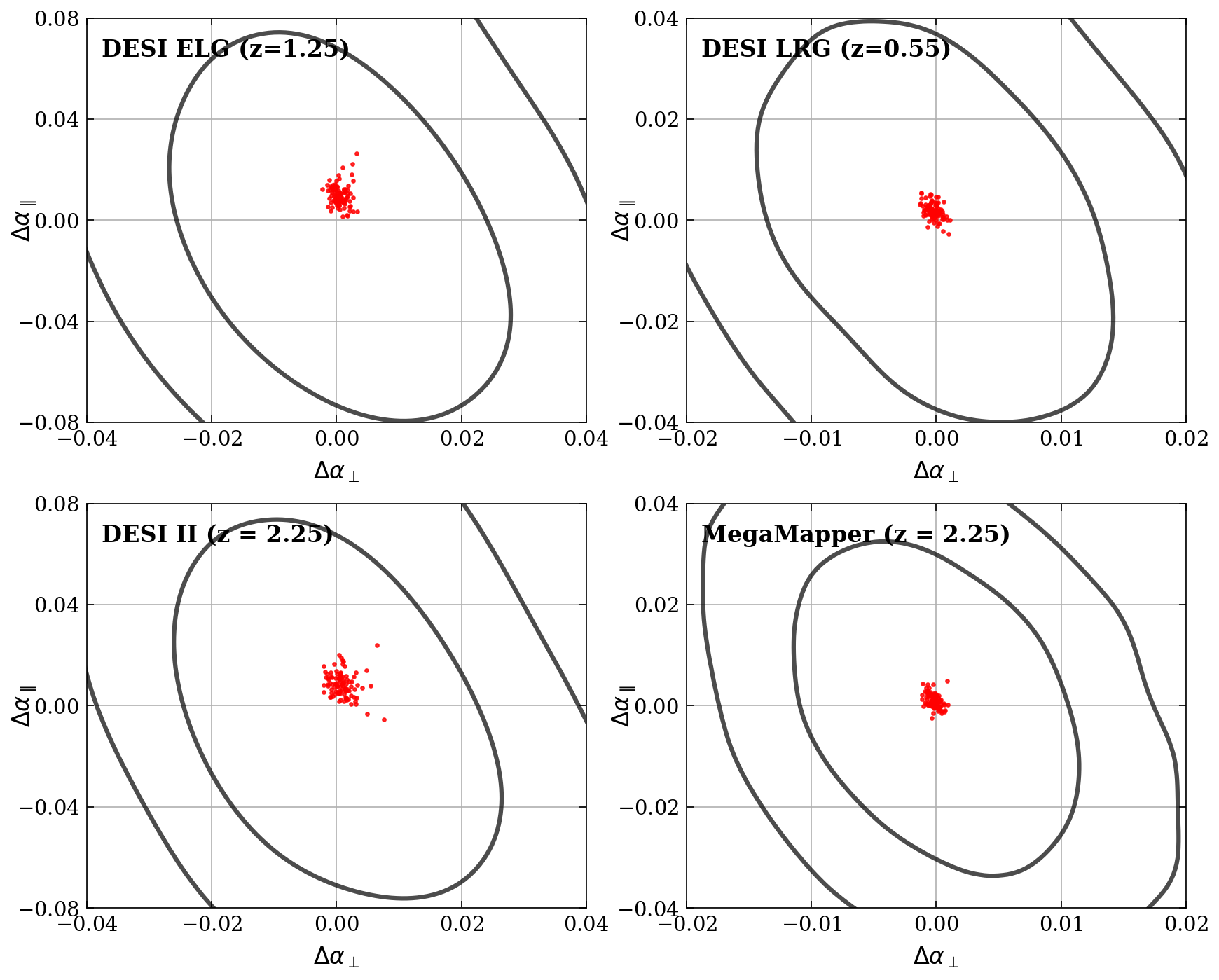}
    \caption{Distribution of the values of $\Delta \alpha_{\parallel}$ and $\Delta \alpha_{\perp}$ (red points) ---  differences between the measured and true values of the alphas--- for mock Horndeski power spectra. The four panels refer to different tracers and survey configurations that we studied, while the black contour in each instance indicates the \textit{typical} 68\% and 95\% credible contours in the $\alpha_{\parallel}$-$\alpha_{\perp}$ plane. This plot illustrates that the biases $\Delta \alpha_{\parallel}$ and $\Delta \alpha_{\perp}$ are well within the typical statistical errors, reflecting the fact that the standard BAO analysis is robust against data assuming modified gravity theory.}
    \label{fig:DESI_II_alpha}
\end{figure}

We next show the full statistics of the recovered $\alpha_{\parallel}$ and $\alpha_{\perp}$ parameters, applied to a sample of approximately 100 Horndeski models\footnote{However, \texttt{EFTCAMB} occasionally generates power spectra that have unstable or unusual behaviors, such as the emergence of a disproportionate peak over some narrow ranges of k values. We excluded five such power spectra from our analysis after going through visual inspection.}. We randomly sample these models from the Horndeski parameters ranges given in Table \ref{tab:parameter} and $\Lambda$CDM parameter values shown in Fig.~\ref{fig:planck_sample}. To study the statistics of the BAO analysis robustness across the sample of models in our analysis, we define the shifts of the alpha parameters relative to their true values,
\begin{equation}
    \begin{aligned}
    \Delta \alpha_{\parallel} &\equiv \alpha_{\parallel, \text{{fit}}} - \alpha_{\parallel, \text{{true}}}\\[0.2cm]
    \Delta \alpha_{\perp} &\equiv \alpha_{\perp, \text{{fit}}} - \alpha_{\perp, \text{{true}}}.
    \end{aligned}
    \label{eq:relative_alphas}
\end{equation}
We focus on the typical values of these shifts relative to typical statistical (measurement) errors in the corresponding alphas. 

Table \ref{tab:delta_alpha} presents the statistics of the shifts in the two alpha parameters derived from realizations conducted on various tracers and galaxy surveys. The mean and standard deviation, $\overline{\Delta\alpha_{\parallel}}$ and $\sigma(\Delta\alpha_{\parallel})$ (and same for $\alpha_{\perp}$), are both found to be very small --- between $10^{-3}$ and $10^{-2}$. Perhaps more usefully, we also show the ratio of the typical deviations in the systematic biases in the alphas relative to their measurement errors, $\sigma(\Delta\alpha_{\parallel})/\sigma(\alpha_{\parallel})$ (and same for $\alpha_{\perp}$).  The typical values of the biases in the best-fit values of the recovered alphas are between 5\% and 8\% of their typical statistical errors. 

We further illustrate our findings in
Fig.~\ref{fig:DESI_II_alpha}, where the red points show the distribution of the biases in the alphas, $\Delta\alpha_{\parallel}$ and $\Delta\alpha_{\perp}$, for the same four survey/tracer/redshift choices as in Fig.~\ref{fig:alpha_plane} and Table \ref{tab:survey}. In each instance, we also show the typical 68\% and 95\% credible contours in the corresponding alphas as the black contour. Given that the fixed fiducial $P(k)$ is used to calculate the covariance matrix, the credible contours look similar across the same survey. Thus, typical contours suffice our illustrative purposes here. We again see that the recovered alphas are close to the true value. The recovered alphas remain well within the statistical error even in the cases with the most extreme biases.

\JP{Furthermore, we tested (in Appendix \ref{appendix:background}) BAO fits to modified-gravity models where both background and perturbations depart from $\Lambda$CDM. 
In this preliminary analysis, we selected a handful of such models that range from being relatively close to agreeing with Planck data, to differing with it at $>$5$\sigma$. We found that the constraints are largely unbiased for these background-and-perturbations-varying models, and most true alphas lie within the fits' 95$\%$ credible contours. Furthermore, the biases in the recovered alpha parameters are small ($<$1$\sigma$) when base cosmological parameters are close (2-4$\sigma$ away) to the fiducial cosmology. These results illustrate the robustness of the compressed BAO analysis even in modified-gravity models with considerable differences between their background evolution and the fiducial cosmology of the template. However, we also find that when both background and perturbations are allowed to vary, the alpha parameters constraints are significantly weaker ($\sim1.5-2$x larger contour sizes compared to models where only the perturbations are non-$\Lambda$CDM), and some models have notably poor constraints on the alpha parameters. While the additional biases when the background is varied may appear to be of concern, it is also the case that variations in the background are already limited and will be getting more so in the future, as the cosmological constraints on the geometrical measures that govern the background (from BAO, type Ia supernova, and other data) get better. We defer for the future a detailed study of compressed BAO fits in modified-gravity models where both background and perturbations are allowed to be significantly different from $\Lambda$CDM.}

Overall, we have found that a compressed analysis of the BAO in Horndeski models returns accurate results for the key parameters $\alpha_{\parallel}$ and $\alpha_{\perp}$, with biases that are well below the statistical error even for a future survey such as MegaMapper.




\section{Conclusions and Discussion}
\label{sec:conclusion}

Standard baryon acoustic oscillation (BAO) analyses compress the clustering information to the location of the BAO peak in the parameters $\alpha$ that are defined separately for the directions perpendicular and parallel to the line of sight and in each redshift bin. These compressed analyses utilize a physically motivated template to isolate the alphas from other information in the 3D power spectrum; the template is pre-computed and typically assumes a fixed cosmological model (e.g.,\ $\Lambda$CDM with concordance values of cosmological parameters). It has been a long-standing question of just how robust this type of analysis is when considering more complex cosmological models. The robustness of this methodology has been tested for some specific departures from $\Lambda$CDM (see the Introduction), but not for a broad class of modified-gravity models. There is some urgency to address this question since a principal goal of the ongoing Stage IV and forthcoming Stage V surveys is precisely to constrain modified gravity with BAO and RSD. 

In this paper, we found that the compressed analysis is robust to a broad Horndeski class of modified-gravity models. Specifically, we have studied models where the perturbations are determined by Horndeski models, with eight additional free parameters that can freely vary, while the background is given by $\Lambda$CDM (with six standard cosmological parameters being varied). We have made use of the \texttt{EFTCAMB} implementation of Horndeski models to carry out the theoretical predictions, and have implemented our own analysis pipeline that follows the standard compressed-analysis approach. For each survey configuration, we studied 100 cosmological models that are not obviously ruled out (that is, that are in $\lesssim$5$\sigma$ tension with Planck 2018 angular power spectrum data). Our results indicate that the biases in the recovered alphas are less than 10\% of the statistical errors even for a Stage-V survey such as MegaMapper. 
\JP{Moreover, we have extended our study by considering (in Appendix \ref{appendix:background}) modified-gravity models with both background and perturbations different significantly from $\Lambda$CDM. Even in that scenario, we found no significant biases ($\lesssim$2$\sigma$) in the recovered alphas. However, this analysis was preliminary and will require further investigation to obtain precise statistical quantification of the fits for these models.} 

Overall, our findings, combined with previous work on other beyond-$\Lambda$CDM models (notably \cite{Bernal2020}), indicate that a compressed analysis based on a $\Lambda$CDM template remains remarkably robust with respect to the choice of the underlying cosmological model.

While we have established the robustness of the compressed analysis of a broad class of modified-gravity models, we did not cover \textit{all} potential modifications of gravity (see \cite{Langlois2019} for a review). For example, one could further study beyond-Horndeski scalar-tensor models which have two additional free functions \cite{Gleyzes2014}, or degenerate higher-order scalar-tensor (DHOST) theories which have higher-order equations of motion \cite{Woodard2015}. One may also be interested in investigating modified-gravity models beyond scalar-tensor theory, such as higher-dimension, tensor-tensor, or tensor-vector-scalar theories \cite{Clifton2012}. Current observational data have not shown statistically significant departures that would favor these models \cite{Noller2019,Sakstein2016, Sugiyama2023}. However, forthcoming galaxy surveys such as DESI, LSST, Euclid and Roman telescopes, and the Stage-V spectroscopic instrument will provide significant improvement in statistical constraints that will make the observational analysis of these models more compelling.

Finally, we note that we have only tested the robustness of the compressed BAO analysis in this paper. This approach is fairly standard and well-established, but there now exist several more general methods that attempt to extract broadband information in the power spectrum beyond the BAO peak locations. These methods include BAO$+$RSD (i.e.\ fitting for f(z)$\sigma_8(z)$, e.g., \cite{neveux2020}), \textit{ShapeFit} \cite{Brieden2021}, as well as direct modeling of the whole power spectrum  \cite{Ivanov2020,Philcox2020,Philcox2022,Troster2020}. These methods are more general than the analysis that simply works off of $\alpha_{\parallel}$ and $\alpha_\perp$ 
and thus offer a greater potential to extract information from high-quality spectroscopic observations. These methods also make use of the broadband power spectrum which, as we have seen (e.g.,\ in our Appendix \ref{appendix:hornps}), is strongly impacted by modified gravity. There is therefore some level of urgency to study the robustness of these more ambitious methods to the underlying cosmological model.


In conclusion, while comparing the performance of different BAO and RSD methods on a wide range of cosmological models remains a priority, we have shown in this paper that the longest-established of such analyses --- the compressed BAO analysis that uses a fixed template --- is very robust in a wide range of modified-gravity cosmological models. 

\section*{Acknowledgements}

We thank Jos\'{e} Luis Bernal and Andreu Font Ribera for helpful discussions and comments on the manuscript, and Nhat Minh Nguyen for useful pointers about the computations with Horndeski models. This research was supported in part through computational resources and services provided by Advanced Research Computing at the University of Michigan, Ann Arbor.
JP, DH, and CA acknowledge support from the Leinweber Center for Theoretical Physics and DOE under contract DE-SC009193. This research was supported in part through computational resources and services provided by Advanced Research Computing at the University of Michigan\footnote{\url{https://arc.umich.edu}} and the University of Michigan Research Computing Package\footnote{\url{https://arc.umich.edu/umrcp}}.

\section*{Data Availability}

The inclusion of a Data Availability Statement is a requirement for articles published in MNRAS. Data Availability Statements provide a standardised format for readers to understand the availability of data underlying the research results described in the article. The statement may refer to original data generated in the course of the study or to third-party data analysed in the article. The statement should describe and provide means of access, where possible, by linking to the data or providing the required accession numbers for the relevant databases or DOIs.



\bibliographystyle{JHEP}

\bibliography{BAO} 




\appendix

\section{Matter power spectrum in Horndeski models}
\label{appendix:hornps}
In this Appendix, we provide an overview of the variations in the matter power spectrum within Horndeski models. First, we show the effect of varying individual Horndeski free parameters on the matter power spectrum, thus complementing the discussion in Sec.~\ref{sec:matterPS} where we illustrated (in Fig.~\ref{fig:pk_final}) the effect of varying simultaneously the Horndeski (perturbation) and $\Lambda$CDM (background) parameters. Next, we present how the compressed analysis best-fitted parameters of the power spectrum in the Horndeski models differ from the $\Lambda$CDM case. 

\begin{figure*}
    \centering
    \resizebox{\textwidth}{!}{%
    \begin{tabular}{@{}cc@{}}
    \includegraphics[width=.25\linewidth,valign=m]{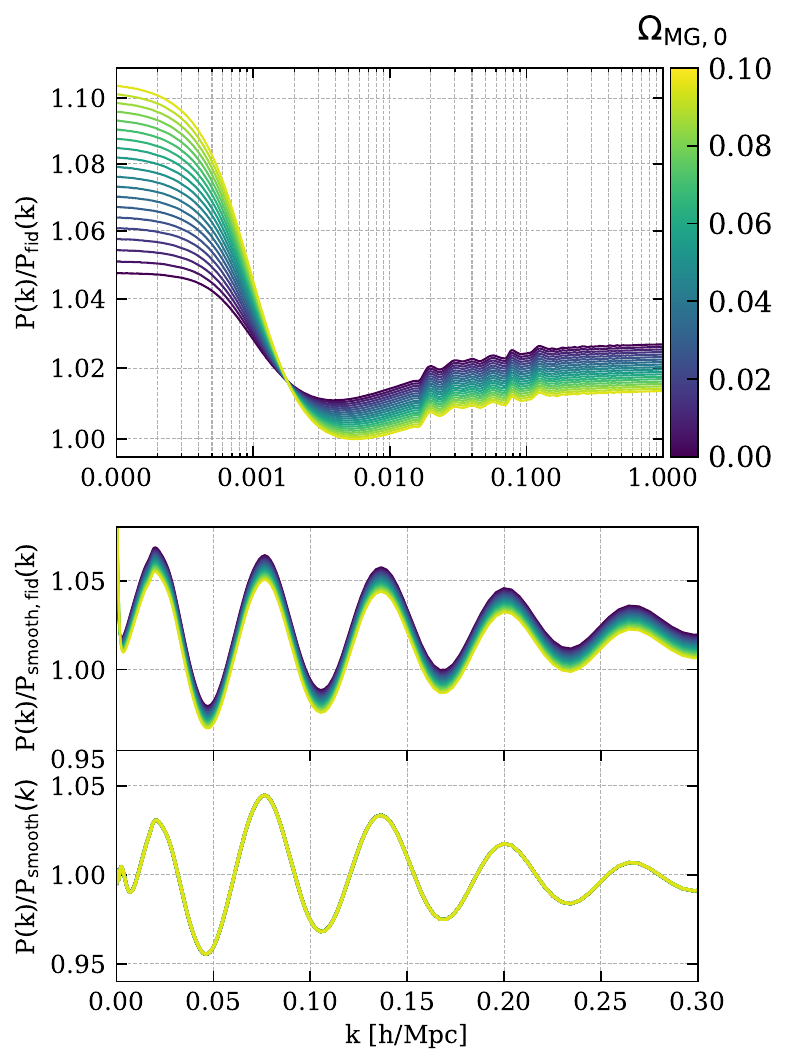} & \includegraphics[width=.25\linewidth,valign=m]{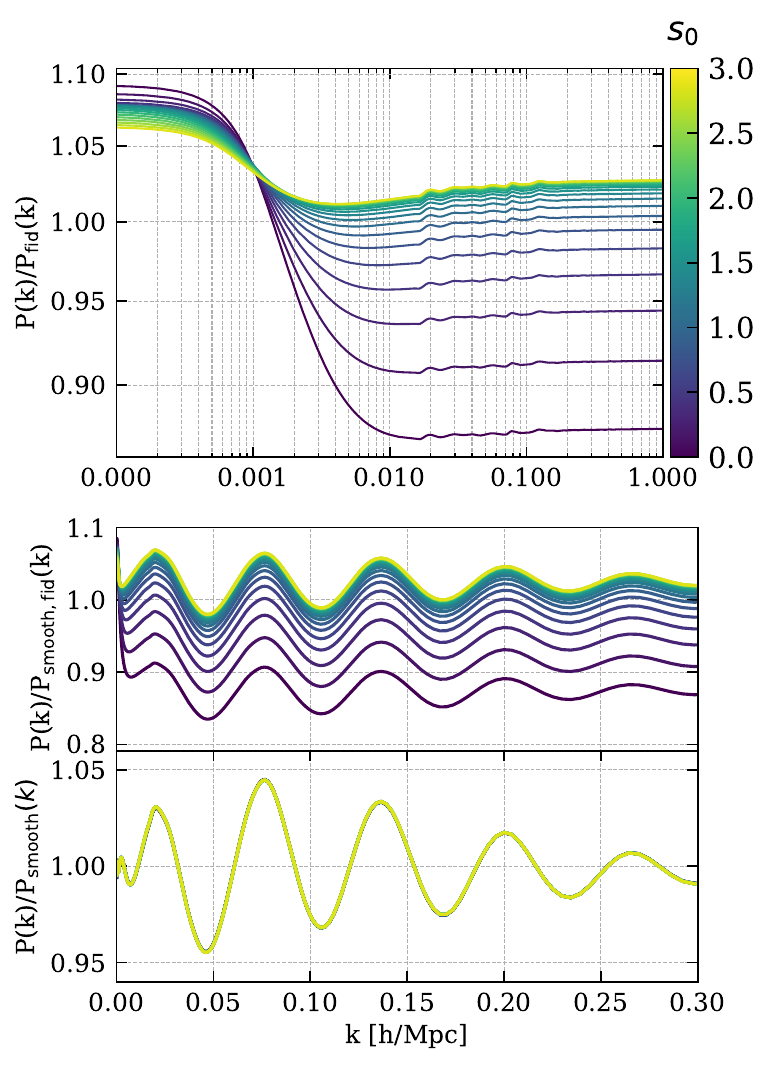} \\
    \end{tabular}
    }
    \caption{ Matter power spectrum in  Horndeski models with varying $\Omega_{\text{MG,0}}$ (left column) and $s_{0}$ (right column). These two parameters control the background evolution of Horndeski models. In each column, the upper panel shows the power spectra divided by fiducial (fixed) power spectrum $P(k)/P_{\text{fid}}(k)$, the middle panel shows the power spectra divided by smoothed component of the fiducial power spectrum, $P(k)/P_{\text{smoooth, fid}}(k)$, while the bottom panel shows the spectrum divided by its own smooth components, $O_{\text{lin}} = P(k)/P_{\text{smoooth}}(k)$. The lower panels therefore characterize the BAO feature; we see that the locations of the BAO do not change when the Horndeski model parameters are varied, but they \textit{do} change once we also vary the base cosmological-model parameters.}
   \label{fig:Omega s0}
\end{figure*}

\begin{figure}[htbp]
     \centering
    \begin{tabular}{ll}
    \includegraphics[width=.33\linewidth,valign=m]{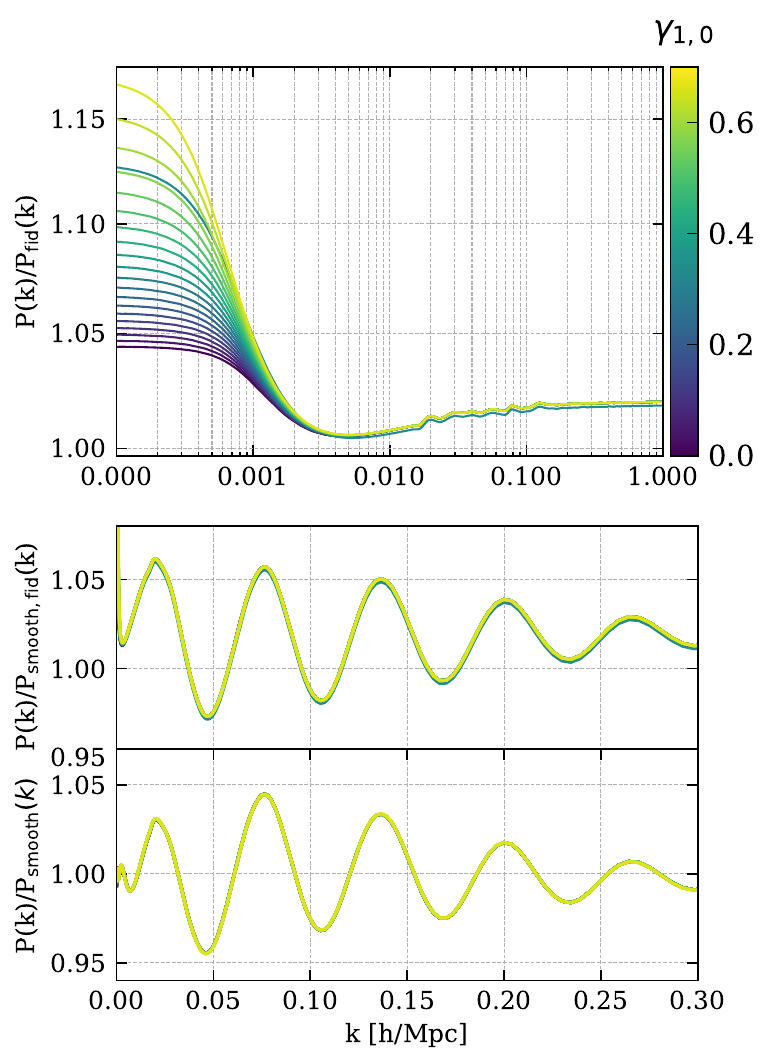} & \includegraphics[width=.33\linewidth,valign=m]{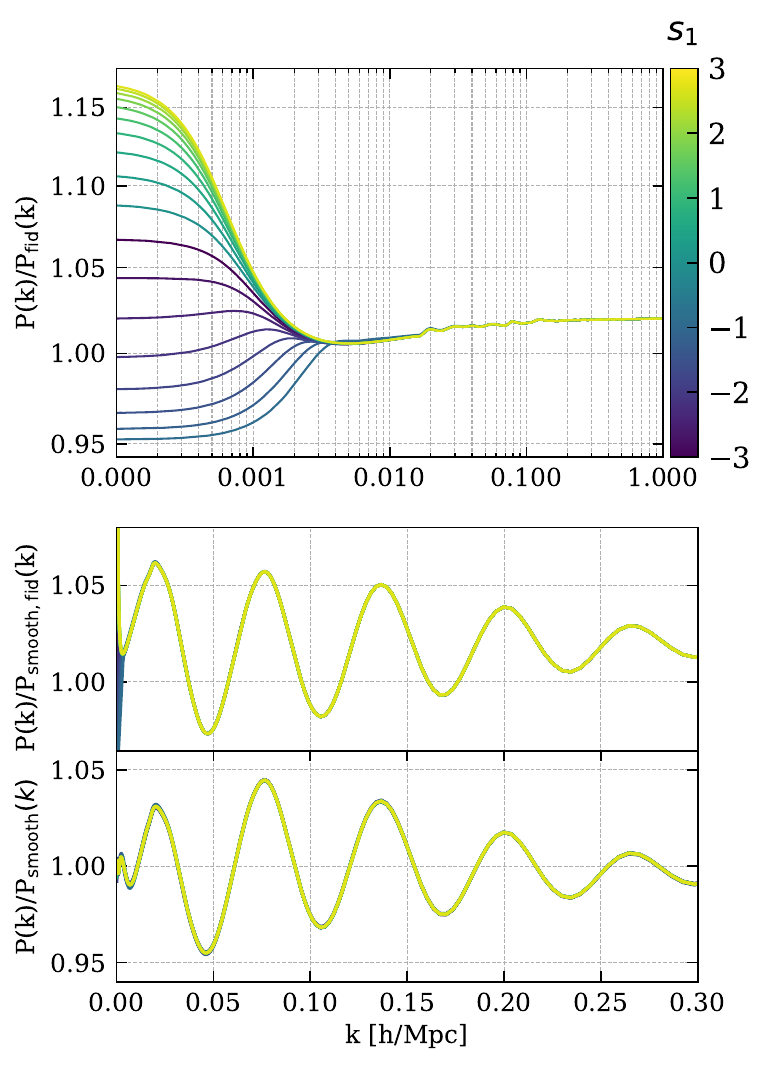} \\
    \includegraphics[width=.33\linewidth,valign=m]{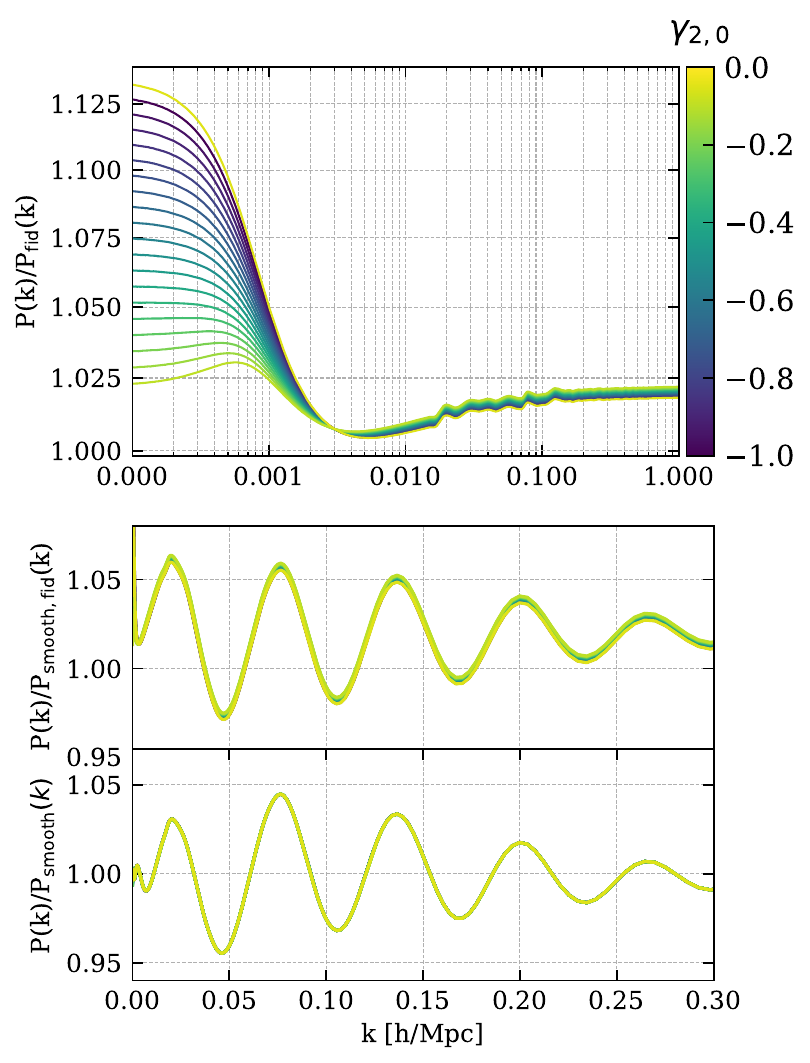} & \includegraphics[width=.33\linewidth,valign=m]{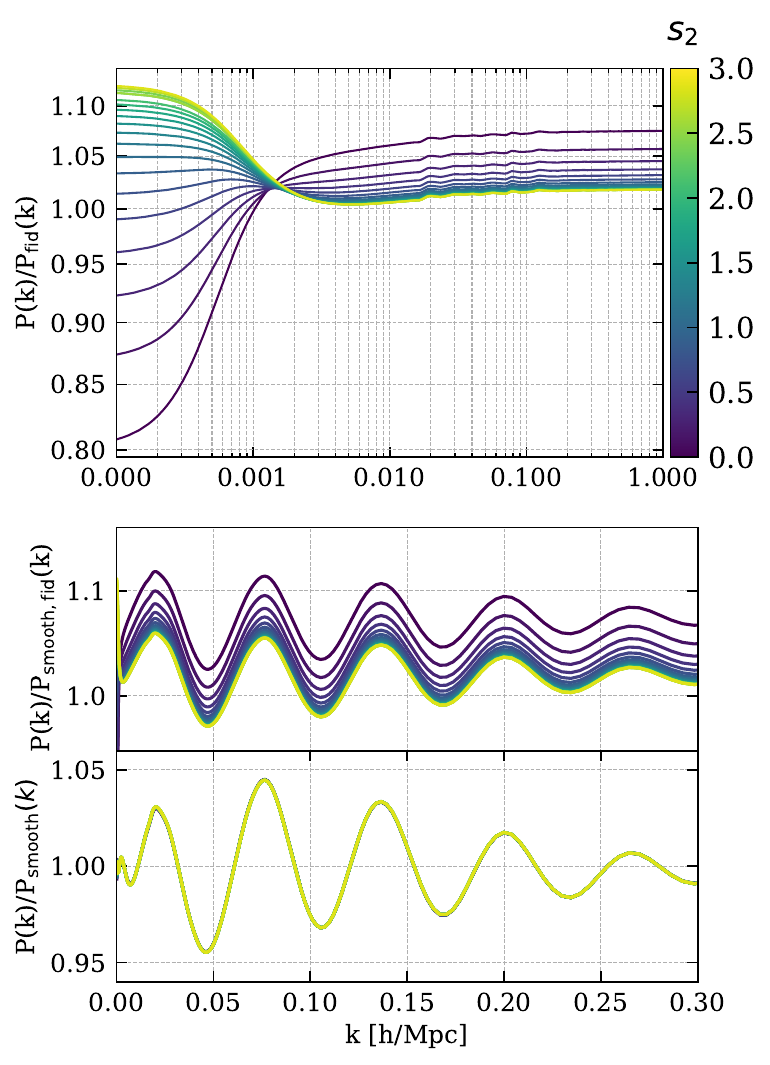} \\
    \includegraphics[width=.33\linewidth,valign=m]{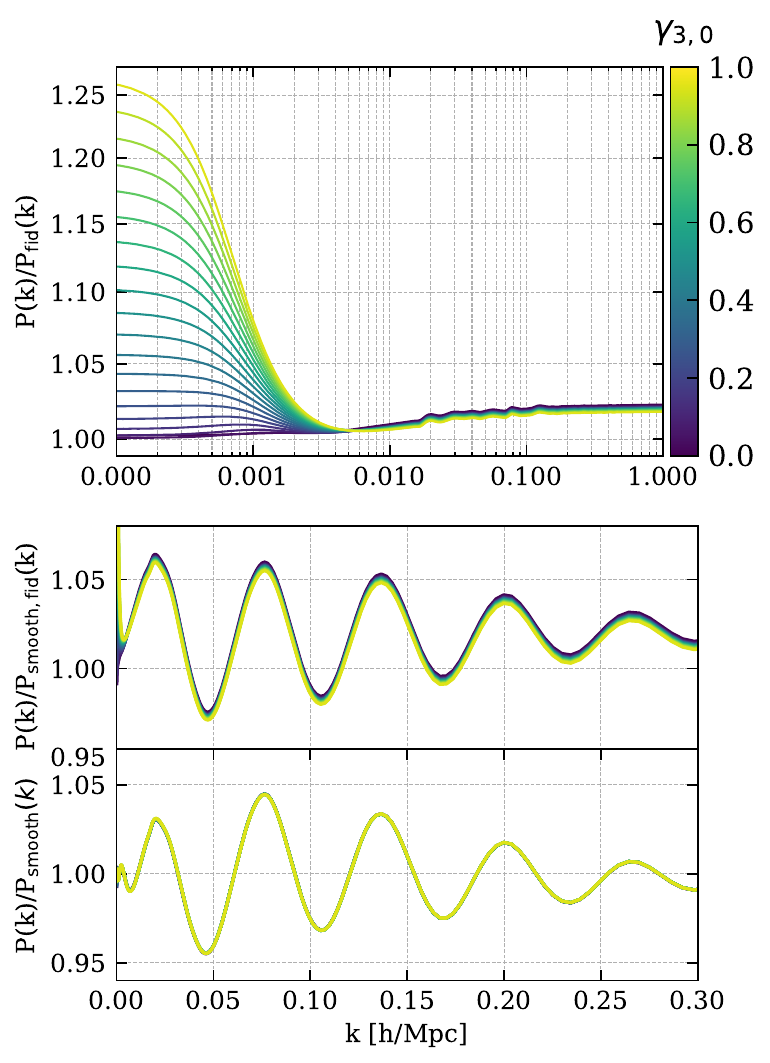} & \includegraphics[width=.33\linewidth,valign=m]{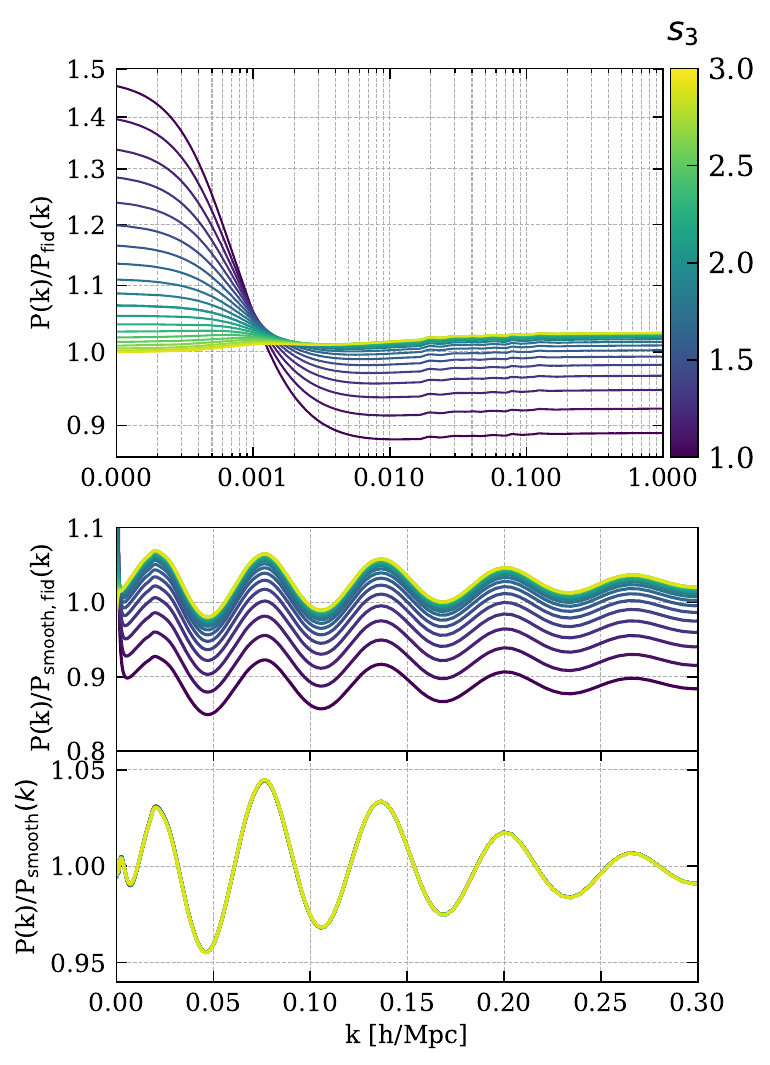} \\
    \end{tabular}
    \caption{The matter power spectrum in the Horndeski models with varying $\gamma_{\text{1,0}}$, $\gamma_{\text{2,0}}$, $\gamma_{\text{3,0}}$, $s_{\text{1}}$, $s_{\text{2}}$, and $s_{\text{3}}$ parameter, which are controlling perturbation evolution in these models. These plots illustrate the modifications in the shapes of the BAO signal by Horndeski models, similar to Fig.~\ref{fig:Omega s0}.
}
    \label{fig:EFTCAMB Paras}
\end{figure}

Fig.~\ref{fig:Omega s0} and Fig.~\ref{fig:EFTCAMB Paras} show the dependence of the power spectrum on the variation of individual Horndeski parameters. Fig.~\ref{fig:Omega s0} shows the effects of $\Omega_{\text{MG}}(a)$, while Fig.~\ref{fig:EFTCAMB Paras} shows that of the $\gamma_i(a)$ (where i goes from 1 to 4). The top panels show the power spectra divided by the fiducial (fixed) power spectrum $P(k)/P_{\text{fid}}(k)$, the middle panels show their ratios to the smooth fiducial power spectrum, $P(k)/P_{\text{smoooth, fid}}(k)$, while the lower panels show the ratios relative to the smoothed portion of each corresponding spectrum, $O_{\text{lin}} = P(k)/P_{\text{smoooth}}(k)$. Note that varying $\gamma_{\text{1,0}}$ and $s_{\text{1}}$ parameters have minimal impact on the BAO feature compared to varying the background.  This outcome aligns with expectations since the contributions from $\gamma_1(a)$ are relatively small when contrasted with observable uncertainties \cite{Frusciante2019}. Thus, we observe that $P(k)/P_{\text{smoooth, fid}}(k)$ is almost unchanged for these two parameters. In particular, we see that varying $\Omega_{\text{MG,0}}$, $s_0$, $s_2$, and $s_3$ mostly affects the shape of the BAO feature. 

We then compare the best-fitted parameters from the analysis applied to one randomly chosen Horndeski model with that applied to the fiducial $\Lambda$CDM model; the results are shown in Table~\ref{tab:bestparas}. These two models have different input background cosmology (that is, the $\Lambda$CDM parameters assumed are different for the two models). The parameters are determined by adopting our analysis procedure on noiseless mock data, and using the least-squares fit implemented in the \texttt{iminuit} tool \cite{James1975dr}, assuming MegaMapper survey with LBG tracers at redshift of $2.25$. The central column lists the best-fitted parameters for the analysis of the Horndeski model, while the right column details the parameters for the analysis of $\Lambda$CDM. 
Note the statistically significant deviation of the parameters $\alpha_{\parallel}$ and $\alpha_{\perp}$ in Horndeski analysis relative to the fiducial values of unity. This deviation is expected as the background cosmological parameters for the Horndeski model power spectrum are different from the fiducial-cosmology values, and lead to shifts in the BAO peak positions. 

One other feature of note seen in Table~\ref{tab:bestparas} is a significant variation in the parameters that model the broadband power spectrum for the Horndeski analysis relative to those in the $\Lambda$CDM analysis. This is especially true for polynomial terms $A_l(k)$ (particularly higher-order terms),  for $\Sigma_{\perp}$ and $\Sigma_{\parallel}$ which model the damping of BAO, and for the galaxy bias ${b_{g}}^2$. These differences between the two analyses are expected, and confirm that the amplitude characteristics of the power spectrum in Horndeski models are markedly distinct from those in $\Lambda$CDM.

Finally, we mention an additional caveat: modified-gravity models may predict different large-scale bulk flows which may in turn affect the BAO. These nonlinear effects manifest as a change in the amplitude of the BAO wiggles and the shape of the BAO feature with respect to those in the $\Lambda$CDM models but are not expected to affect the BAO positions. Such effects are modeled by means of IR resummation with time-sliced perturbation theory \cite{Blas2016} in the direct modeling approach \cite{Ivanov2020}, and this modeling is expected to be accurate if higher-order calculations are included. However, the compressed analysis, which employs an exponential suppression term to model the BAO wiggles amplitude (as seen in Eq.~\ref{eq:aniPS}), may not be sufficiently accurate to model the bulk-flow effects in modified gravity models. Our results indicate that adopting the exponential-suppression term in modified gravity models remains sufficiently accurate, as it does not bias the estimation of $\alpha_{\parallel}$ and $\alpha_{\perp}$. Nonetheless, determining accurate analytical expressions for the exponential suppression term in power spectra under modified gravity models needs future investigation.

\begin{table}
    \centering
    \caption{Comparison of BAO compressed analysis best-fit parameters in the fiducial $\Lambda$CDM model and in a Horndeski model. Here, the set of parameters is for the anisotropic power spectrum fitting template.}
    \begin{tabular}{|c|r@{$\pm$}l|r@{$\pm$}l|}
        \hline
        \textbf{Parameter} & \multicolumn{2}{c|}{\textbf{Values (Horndeski)}} & \multicolumn{2}{c|}{\textbf{Values ($\Lambda$CDM)}} \\
        \hline
        $\alpha_{\parallel}$ & 1.050 & 0.004 & 1.000 & 0.007 \\
        $\alpha_{\perp}$ & 0.943 & 0.010 & 1.000 & 0.012 \\
        $a_{0,1}$ & 0.123 & 0.005 & 0.000 & 0.005 \\
        $a_{0,2}$ & -13.57 & 0.423 & -0.010 & 0.350 \\
        $a_{0,3}$ & 110.3 & 2.013 & 0.240 & 2.469 \\
        $a_{0,4}$ & 1033 & 3.761 & 1264 & 10.41 \\
        $a_{0,5}$ & 69.72 & 13.11 & 0.674 & 30.05 \\
        $\Sigma_{\text{FoG}}$ & 8.363 & 0.110 & 9.998 & 0.166 \\
        ${b_{g}}^2$ & 10.35 & 0.128 & 4.489 & 0.065 \\
        $\beta$ & 0.291 & 0.012 & 0.162 & 0.018 \\
        $a_{2,1}$ & 0.050 & 0.011 & 0.000 & 0.011 \\
        $a_{2,2}$ & -4.982 & 0.760 & -0.013 & 0.722 \\
        $a_{2,3}$ & -64.36 & 3.722 & -0.934 & 4.951 \\
        $a_{2,4}$ & 131.1 & 4.417 & -0.510 & 22.30 \\
        $a_{2,5}$ & 112.2 & 42.23 & -10.02 & 64.13 \\
        $\Sigma_{\perp}$ & 7.662 & 0.867 & 3.988 & 1.226 \\
        $\Sigma_{\parallel}$ & 18.56 & 1.922 & 8.015 & 2.300 \\
        \hline
    \end{tabular}
\label{tab:bestparas}
\end{table}

\section{Comparing Power Spectrum Smoothing Methods: Direct Interpolation vs. Indirect Approaches}
\label{appendix:smoothb}
\begin{figure}	
\centering
\includegraphics[width=\columnwidth]{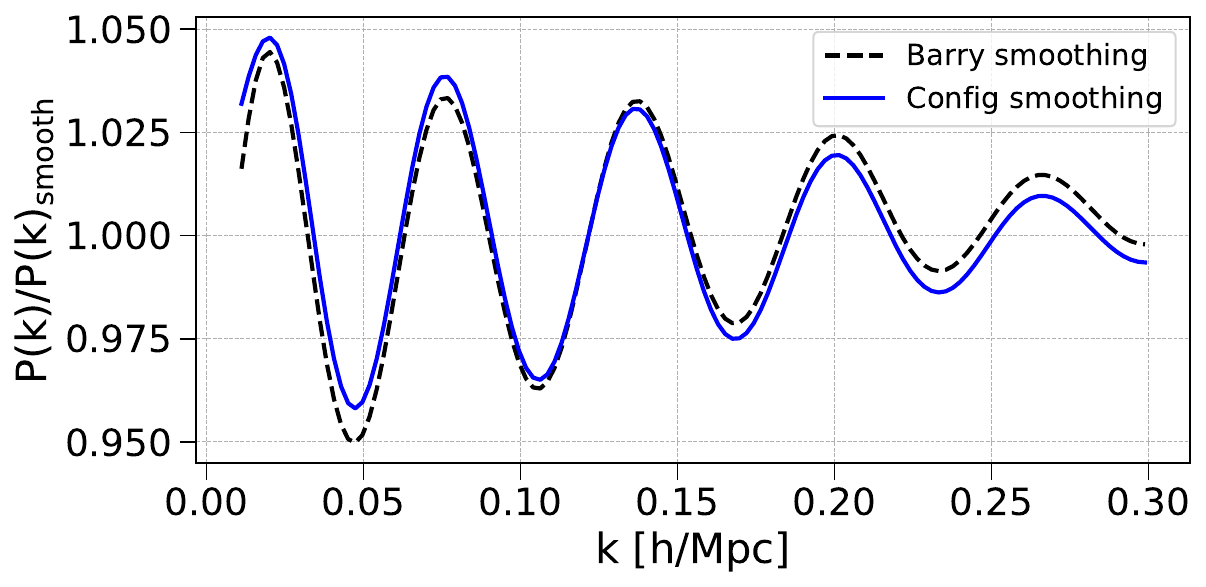}
     \caption{Comparison of power spectra with different smoothing methods. This plot shows the ratio of the power spectrum, $P(k)$, to its smoothed versions using the smoothing method used in Barry framework \cite{Hinton2017} (solid line) and the indirect smoothing method (dashed line).}
    \label{fig:smooth comp}
\end{figure}

Here we investigate the differences between the methods used to extract smooth components of the power spectrum in \texttt{Barry} code \cite{Hinton2017} and those adopted by \cite{Bernal2020}. \texttt{Barry} employs direct interpolation on the power spectrum to achieve smoothing; specifically, it employs polynomial functions to dewiggle the power spectrum.  On the other hand, the \cite{Bernal2020} approach involves a conversion to the configuration space for smoothing, followed by reconversion to the Fourier space; this procedure also smoothes the power spectrum. Fig.~\ref{fig:smooth comp} visually contrasts these two methodologies. In the figure, we define the difference between the power spectra, $\Delta P(k)$, as:
\begin{equation}
\Delta P(k) = |P_{\text{sm, config}}(k) - P_{\text{sm, barry}}(k)|.
\end{equation}
Here, $P_{\text{sm, config}}(k)$ denotes the smoothed power spectrum obtained from the configuration-space method presented in \cite{Bernal2020}, while $P_{\text{sm, barry}}(k)$ corresponds to the spectrum derived using the Barry code.

To test the effects of these two smoothing methods, we apply them to the BAO compressed analysis for a single Horndeski model, with all other choices (e.g.,\ survey specifications) being the same. The smoothing method in the \texttt{Barry} code reports $\alpha_{\parallel} = $ 0.986 $\pm$ 0.048 and $\alpha_{\perp} = $ 0.974 $\pm$ 0.010. In comparison, the smoothing method in the configuration space reports $\alpha_{\parallel} =$ 0.985 $\pm$ 0.028 and $\alpha_{\perp} = $ 0.975 $\pm$ 0.010. The differences between the best-fitted $\alpha_{\parallel}$ and $\alpha_{\perp}$ are less than $0.1\%$ for these two smoothing methods, well below the statistical errors in the alphas. While the resulting peak locations agree extremely well between the two methods, we also see in Fig.~\ref{fig:smooth comp} that the resulting amplitude of $P(k)/P(k)_{\rm smooth}$, which is of less importance of the BAO analysis, also agrees well, with a typical difference around $0.5 \%$ between the two smoothing methods. 

Therefore, we have shown that the \texttt{Barry} and configuration-space smoothing methods agree very well, and the choice of which one to pick will have a negligible effect on the final results.


\section{Exploring modified-gravity models significantly deviating from $\Lambda$CDM in both background and perturbations}
\label{appendix:background}

\JP{Here we perform tests of the compressed BAO analysis on modified-gravity models in which both background and perturbations are allowed to depart from $\Lambda$CDM. We performed the compressed BAO analysis on ten models randomly selected according to the procedure in Section \ref{sec:procedure}, so that the models are in the $<$5$\sigma$ deviation from Planck's best-fit cosmology in the multi-dimensional parameter space. Two of the model fits had very poor constraints on the alpha parameters, and we omitted them from further analysis\footnote{The models with a poor fit to the alphas do nominally also sometimes have a larger deviation from the true alphas. However, we do not think that these bad fits are a concern: a poor fit to the alphas is a sign that the template used is unsuitable, so in those cases, the compressed BAO analysis would have to be redone with a better template. We are therefore justified in dropping these cases from our analysis.}. Moreover, we selected five more "extreme" models that are 5-6$\sigma$ away from Planck's best fit; of these, three models had good fits on the alphas and we consider them further. For these preliminary tests of models where both the background and perturbations differ from $\Lambda$CDM, we only considered the MegaMapper survey settings.}

\begin{figure}	
\centering
\includegraphics[width=0.75\columnwidth]{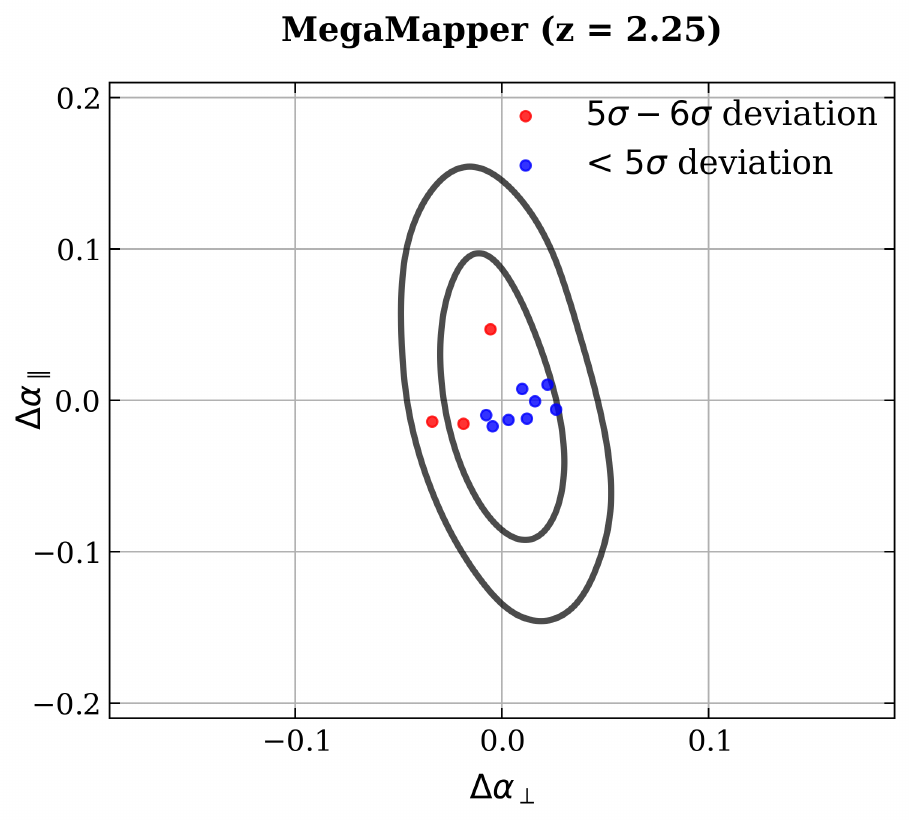}
     \caption{\JP{Distribution of the values of $\Delta \alpha_{\parallel}$ and $\Delta \alpha_{\perp}$, representing differences between the best-fit values of the alphas, for a total of 11 randomly selected modified-gravity models where both background and perturbations differ from $\Lambda$CDM.  The eight blue points are modified-gravity models that are $<$5$\sigma$ away from Planck's best-fit cosmology, while the three red points represent "extreme" models that deviate from Planck by 5-6$\sigma$. The black contours indicate the typical 68$\%$ and 95$\%$ credible contours in the $\alpha_{\parallel}$-$\alpha_{\perp}$ plane. Note that 9 out of 11 models have biases in the alphas that are less than 1$\sigma$. We assume the MegaMapper survey settings here.}}
    \label{fig:background}
\end{figure}

\JP{Fig.~\ref{fig:background} shows the fits to the alphas for these 11 modified-gravity models. The eight models with $<$5$\sigma$ deviation from Planck are shown as blue points, while the three models with 5-6$\sigma$ deviation are shown as red points. We find that biases in the recovered alpha parameters are all below 2$\sigma$, and in 9 out of 11 cases, they are less than 1$\sigma$. The performance of the compressed BAO analysis therefore remains excellent even for these modified-gravity models with more freedom, at least based on this preliminary analysis with limited statistics.
}


\JP{To accommodate potential cases where the underlying model significantly deviates from $\Lambda$CDM, the performance of the BAO compressed analysis can also be improved by undertaking additional steps (not adopted in this paper). For example, one can extract information from BAO without relying on a precomputed template (e.g., \cite{Colas2020,Philcox2020}). In this case, at each step of the analysis, a linear power spectrum and high-order loop corrections are calculated to infer parameters of interest (instead of using a fixed template).
Alternatively, one can add new degrees of freedom to the fiducial cosmology in the template to account for the possibility of modified gravity (by, for example,  adding extra nuisance parameters to capture the impact of $\Omega_{\text{MG}}(t)$ and $\gamma_i(a)$). Investigating these possibilities will enable further overall robustness of the BAO methodology.
} 

\end{document}